\documentclass{aastex}
\usepackage{emulateapj5,apjfonts}

\newcommand{\gcm}{\mbox{g cm$^{-3}$}}

\newcommand{\pcmsq}{\mbox{cm$^{-2}$}}
\newcommand{\ergsec}{\mbox{ergs s$^{-1}$}}
\newcommand{\ergcms}{\mbox{ergs cm$^{-2}$ s$^{-1}$}}
\newcommand{\kmsec}{\mbox{km s$^{-1}$}}
\newcommand{\Lx}{\mbox{$L_{\rm x}$}}
\newcommand{\Fx}{\mbox{$f_{\rm x}$}}

\newcommand{\Msun}{\mbox{M$_\odot$}}
\newcommand{\ml}{\mbox{\Msun\ yr$^{-1}$}}
\newcommand{\hr}{\mbox{$^{\rm h}$}}
\newcommand{\minute}{\mbox{$^{\rm m}$}}

\newcommand{\chandra}{\textit{Chandra}}

\newcommand{\mekal}{\textsc{MEKAL}}
\newcommand{\vmek}{\textsc{VMEKAL}}
\newcommand{\xspec}{\textsc{XSPEC}}

\shorttitle{SNe 1999em and 1998S}
\shortauthors{Pooley et al.}
\slugcomment{Accepted by the {\it Astrophysical Journal}}

\begin{document}

\submitted{Accepted by the {\it Astrophysical Journal}}

\title{X-ray, Optical, and Radio Observations of the Type II
Supernovae 1999em and 1998S}

\author{David Pooley\altaffilmark{1},
Walter~H.~G.~Lewin\altaffilmark{1}, Derek W. Fox\altaffilmark{1,2}, Jon
M.~Miller\altaffilmark{1}, 
Christina K. Lacey\altaffilmark{3}, 
Schuyler D. Van Dyk\altaffilmark{4}, 
Kurt W. Weiler\altaffilmark{5}, 
Richard A. Sramek\altaffilmark{6},
Alexei V. Filippenko\altaffilmark{7}, 
Douglas C. Leonard\altaffilmark{7,8},
Stefan Immler\altaffilmark{9}, 
Roger A. Chevalier\altaffilmark{10},
Andrew C. Fabian\altaffilmark{11}, 
Claes Fransson\altaffilmark{12},
Ken'ichi Nomoto\altaffilmark{13}}

\altaffiltext{1}{Center for Space Research and Department of Physics,
Massachusetts Institute of Technology, Cambridge, MA  02139-4307;
dave@mit.edu, lewin@space.mit.edu, derekfox@space.mit.edu, jmm@space.mit.edu} 
\altaffiltext{2}{Present Address: Astronomy Department, California
Institute of Technology, Mail Code 105-24, Pasadena, CA 91125} 
\altaffiltext{3}{Department of Physics and Astronomy, University of
South Carolina, Columbia, SC  29208; lacey@sc.edu}
\altaffiltext{4}{Infrared Processing and Analysis Center, California
Institute of Technology, Mail
Code 100-22, Pasadena, CA  91125; vandyk@ipac.caltech.edu} 
\altaffiltext{5}{Naval Research Laboratory, Code 7213, Washington DC
20375-5320; weiler@rsd.nrl.navy.mil}
\altaffiltext{6}{National Radio Astronomy Observatory, Socorro, NM
87801; dsramek@nrao.edu} 
\altaffiltext{7}{Department of Astronomy, University of California,
Berkeley, CA 94720-3411; alex@astron.berkeley.edu}
\altaffiltext{8}{Present address: Five College Astronomy Department, University of Massachusetts,
Amherst, MA  01003-9305; leonard@nova.astro.umass.edu}
\altaffiltext{9}{Astronomy Department, University of Massachusetts,
Amherst, MA  01003-9305; immler@astro.umass.edu}
\altaffiltext{10}{Department of Astronomy, University of Virginia,
Charlottesville, VA  22903; rac5x@virginia.edu}
\altaffiltext{11}{Institute of Astronomy, Madingley Road, Cambridge,
CB3 OHA, England, UK; acf@ast.cam.ac.uk}
\altaffiltext{12}{Stockholm Observatory, SE-133 36 Saltsj\"{o}baden,
Sweden; claes@astro.su.se}
\altaffiltext{13}{Department of Astronomy and Research Center for the
Early Universe, School of Science, University of Tokyo,
Bunkyo-ku, Tokyo 113-0033, Tokyo, Japan; nomoto@astron.s.u-tokyo.ac.jp}

\begin{abstract}
Observations of the Type II-P (plateau) Supernova (SN) 1999em and Type
IIn (narrow emission line) SN~1998S have enabled estimation of the
profile of the SN ejecta, the structure of the circumstellar medium
(CSM) established by the pre-SN stellar wind, and the nature of the
shock interaction.  SN~1999em is the first Type II-P detected at both
X-ray and radio wavelengths.  It is the least radio luminous and one
of the least X-ray luminous SNe ever detected (except for the unusual
and very close SN~1987A).  The \chandra\ X-ray data indicate
non-radiative interaction of SN ejecta with a power-law density
profile ($\rho \propto r^{-n}$ with $n\sim 7$) for a pre-SN wind with
a low mass-loss rate of $\sim 2 \times 10^{-6}$~\ml\ for a wind
velocity of 10~\kmsec, in agreement with radio mass-loss rate
estimates.  The \chandra\ data show an unexpected, temporary rise in
the 0.4--2.0~keV X-ray flux at $\sim$100 days after explosion.
SN~1998S, at an age of $>$3 years, is still bright in X-rays and is
increasing in flux density at cm radio wavelengths.  Spectral fits to
the \chandra\ data show that many heavy elements (Ne, Al, Si, S, Ar,
and Fe) are overabundant with respect to solar values.  We compare the
observed elemental abundances and abundance ratios to theoretical
calculations and find that our data are consistent with a progenitor
mass of approximately 15--20~\Msun\ if the heavy element ejecta are
radially mixed out to a high velocity.  If the X-ray emission is from
the reverse shock wave region, the supernova density profile must be
moderately flat at a velocity $\sim 10^4$~\kmsec, the shock front is
non-radiative at the time of the observations, and the mass-loss rate
is 1--2$\times 10^{-4}$~\ml\ for a pre-supernova wind velocity of
$10$~\kmsec.  This result is also supported by modeling of the radio
emission which implies that SN~1998S is surrounded by a clumpy or
filamentary CSM established by a high mass-loss rate, $\sim 2 \times
10^{-4}$~\ml, from the pre-supernova star.
\end{abstract}


\section{Introduction}
To date, 14 supernovae (SNe) have been detected in X-rays in the near
aftermath (days to years) of their explosions\footnote{A complete list
of X-ray SNe and references can be found at
\url{http://xray.astro.umass.edu/sne.html}.}: SNe~1978K
\citep{petre94,schlegel95,schlegel96,schlegel99a}, 1979C
\citep{immler98a,kaaret01,ray01}, 1980K
\citep{can82,schlegel94,schlegel95}, 1986J
\citep{houck98,schlegel95,chug93,bregpil92}, 1987A
\citep{bur00,dennerl01,gore94,hasinger96}, 1988Z \citep{fab96}, 1993J
\citep{immler01,suzno95,zim94}, 1994I \citep{immler98b,immler02a},
1994W \citep{schlegel99b}, 1995N \citep{fox00}, 1998bw (aka
GRB~980425; Pian et al.\ 1999; Pian et al.\ 2000), 1999em
\citep{fox99,schlegel01a}, and 1999gi \citep{schlegel01b}.  We present
here results from \chandra\ observations of SN~1999em and SN~1998S.
In general, the high X-ray luminosities of these 14 SNe ($\Lx\sim
10^{38}$--10$^{41}$ \ergsec) dominate the total radiative output of
the SNe starting at an age of about one year.  This soft X-ray ($\la
10$~keV) emission is most convincingly explained as thermal radiation
from the ``reverse shock'' region that forms within the expanding SN
ejecta as it interacts with the dense stellar wind of the progenitor
star.

The interaction of a spherically symmetric SN shock and a smooth
circumstellar medium (CSM) has been calculated in detail
\citep{chevfran94,suzno95}.  As the SN ``outgoing'' shock emerges from
the star, its characteristic velocity is $\sim$10$^4$~\kmsec, and the
density distribution in the outer parts of the ejecta can be
approximated by a power-law in radius, $\rho \propto r^{-n}$, with
$7\la n\la 20$.  The outgoing shock propagates into a dense CSM formed
by the pre-SN stellar wind.  This wind is slow ($v_w \sim 10~\kmsec$)
and was formed by a pre-supernova mass-loss rate of $\dot{M} \sim
10^{-4}$ to $10^{-6}$~\ml.  The CSM density for such a wind decreases
as the square of the radius ($\rho = \dot{M} / 4 \pi r^2 v_w$). The
collision between the stellar ejecta and the CSM produces a
``reverse'' shock, which travels outward at $\sim$10$^3$~\kmsec\
slower than the fastest ejecta. Interaction between the outgoing shock
and the CSM produces a very hot shell ($\sim$10$^9$ K) while the
reverse shock/ejecta interaction produces a denser, cooler shell
($\sim$10$^7$ K) with much higher emission measure and is where most
of the observable X-ray emission arises.  If either the CSM density or
$n$ is high, the reverse shock is radiative, resulting in a dense,
partly-absorbing shell between the two shocks.

\citet{chev82} proposed that the outgoing shock from the SN explosion
generates the relativistic electrons and enhanced magnetic field
necessary for synchrotron radio emission.  The ionized CSM initially absorbs
most of this emission, except in cases where synchrotron
self-absorption dominates, as may have been the case in
SN~1993J \citep{chev98,franbjor98}.  However, as the shock passes
rapidly outward through the CSM, progressively less ionized material
is left between the shock and the observer, and absorption decreases
rapidly. The observed radio flux density rises accordingly.  At
the same time, emission from the shock region is decreasing slowly as
the shock expands so that, when radio absorption has become
negligible, the radio light curve follows this decline.  Observational
evidence also exists at optical wavelengths for this interaction of SN
shocks with the winds from pre-supernova mass loss \citep{fil97}.

All known radio supernovae appear to share common properties of (i)
non-thermal synchrotron emission with a high brightness temperature,
(ii) a decrease in absorption with time, resulting in a smooth, rapid
turn-on first at shorter wavelengths and later at longer wavelengths,
(iii) a power-law decline of the emission flux density with time at
each wavelength after maximum flux density (absorption $\tau \approx
1$) is reached at that wavelength, and, (iv) a final, asymptotic
approach of the spectral index $\alpha$ to an optically thin,
non-thermal, constant negative value \citep{weiler86}.  These
properties are consistent with the Chevalier model.

The signatures of circumstellar interaction in the radio, optical, and
X-ray regimes have been found for a number of Type II SNe such as the Type
II-L SN~1979C \citep{fessmat93,immler98a,weiler86,weiler91} and the Type
II-L SN~1980K \citep{can82,fesbeck90,leib91,weiler86,weiler92}. The
Type IIn subclass has peculiar optical characteristics: narrow
H$\alpha$ emission superposed on a broad base; lack of P Cygni
absorption-line profiles; a strong blue continuum; and slow evolution
\citep{schlegel90,fil97}.  The narrow optical lines are clear evidence for
dense circumstellar gas --- they probably arise from the reprocessing
of the X-ray emission --- and are another significant means by which
the shock radiatively cools.  The best recent examples of Type IIn SNe
include SNe~1978K, 
1986J, 
1988Z, 
1994W 
(which may have been a peculiar Type II-P, \citealt{sol98}),
1995N, 
1997eg, 
and 1998S. 

SN~1999em was optically discovered on 1999 October 29 in NGC~1637,
approximately 15\arcsec\ west and 17\arcsec\ south of the galactic
nucleus \citep{li99}.  The spectrum on 1999 October 29 shows a high
temperature \citep{baron00}, indicating that the supernova was
discovered at an early phase.  We take 1999 October 28 as the date of the
explosion.  SN~1999em reached a peak brightness of $m_V \approx 13.8$
mag on about day~4, and remained in the ``plateau phase,'' an enduring
period of nearly constant V-band brightness, until about 95 days after
discovery \citep{leon01b}.  At an estimated distance of 7.8~Mpc
(Sohn \& Davidge 1998; see also Hamuy et al.\ 2001 who find a distance
of 7.5~Mpc to SN~1999em based on the Expanding Photosphere Method), this
is the closest Type II-P supernova yet observed, and it is the only
Type II-P observed in both X-rays and radio.  Its maximum brightness
of $M_V \approx - 15.75$ marks it as one of the optically least luminous SNe
II-P \citep{patat94}.  \citet{smartt01} recently derived a mass of
$12\pm1$~\Msun\ for the progenitor of SN~1999em.

On 1998 March 3, SN~1998S was optically discovered 16\arcsec\ west and
46\arcsec\ south of the nucleus of NGC~3877 \citep{li98}.  Its
spectrum quickly revealed it to be a peculiar Type IIn, with narrow H
and He emission lines superimposed on a blue continuum
\citep{filmor98} along with strong N~III, C~II, C~III, and C~IV
emission \citep{gar98,liu00} similar to the ``Wolf-Rayet'' features
seen in the early observations of SN~1983K \citep{niem85}.  SN~1998S
reached a peak brightness of $m_V\approx12$ \citep{gran98,gar98b}.
For a distance of 17~Mpc \citep{tully88}, the intrinsic brightness of
$M_V\approx-18.8$ marks SN~1998S as one of the optically brightest
Type II SNe \citep{patat93}.

\section{Observations}
\subsection{SN~1999em}
\subsubsection{X-ray}
The trigger for our first \chandra\ observation was the optical
discovery of a SN within 10~Mpc.  The extremely rapid response of the
\chandra\ staff allowed us to catch SN~1999em three days after
discovery.  We observed SN~1999em on 1999 November 1, November 13,
December 16, 2000 February 7, October 30, 2001 March 9, and July 22 (days 4,
16, 49, 102, 368, 495, and 630 from reference).  A summary of the
observations is given in Table~\ref{tab:obs99em-xray}, and
\citet{schlegel01a} also discusses the first three.  All observations
were performed by the Advanced CCD Imaging Spectrometer (ACIS) with
the telescope aimpoint on the back-side illuminated S3 chip, which
offers increased sensitivity to low energy X-rays over the front-side
illuminated chips.  The first two observations used a 1/2 subarray
mode (only half of the CCD was read out), while the rest used the full
CCD.  The data were taken in timed-exposure (TE) mode using the
standard integration time of 3.2~sec per frame and telemetered to the
ground in very faint (VF) mode for the first two observations and
faint (F) mode for the rest.  In VF mode, the telemetered data contain
the values of 5 $\times$ 5 pixel islands centered on each event, while
in F mode the islands are 3 $\times$ 3 pixels.

\subsubsection{Radio}
We observed SN~1999em over 30 times with the Very Large Array
(VLA)\footnote{The VLA is operated by the NRAO of the AUI under a
cooperative agreement with the NSF.} at frequencies ranging from 43.3
GHz to 1.5 GHz and ages from 3 to 454 days after the discovery date of
1999 October 29 \citep{lacey01}.  Except for six weak but significant
($>3 \sigma$) detections at mid-cm wavelengths (1.5 -- 8.4 GHz) at
ages of between 33 and 69 days, all measurements yielded only upper
limits.

\subsubsection{Optical}
Extensive optical ground-based observations were made of SN~1999em;
these include {\it UBVRI} photometry, long-term spectroscopy, and
spectropolarimetry \citep{leon01a}.  These data were used to obtain a
distance with the Expanding Photosphere Method \citep{leon01b}.  Both
the photometry and spectroscopy suggest that this was a relatively
normal Type II-P event.  We note, however, that SN 1999em was about 1
magnitude fainter in the V-band during the plateau than the average of
the 8 SNe II-P with published photometry and previously derived EPM
distances, and it may have also had a somewhat unusual color evolution
\citep{leon01b}.

Spectropolarimetry of SN~1999em was obtained on 1999 November 5,
December 8, and December 17 (7, 40, and 49 days after discovery,
respectively, while it was still in its plateau phase) with the Kast
double spectrograph \citep{millstone93} with polarimeter at the
Cassegrain focus of the Shane 3-m telescope at Lick
Observatory. Similarly, on 2000 April 5 and 9 (159 and 163 days after
discovery, long after the plateau). The Low-Resolution Imaging
Spectrometer \citep{oke95} was used in polarimetry mode
\citep{cohen96} at the Cassegrain focus of the Keck-I 10-m telescope.
\citet{leon01a} discuss the details of the polarimetric observations
and reductions.

Total flux spectra \footnote{These spectra are produced from all
available source photons, as opposed to the ``polarized flux spectra''
which refer only to the net polarization as a function of wavelength.}
were obtained at many epochs, primarily with the Nickel 1-m and Shane
3-m reflectors at Lick Observatory. {\it UBVRI} photometry was
conducted with the 0.8-m Katzman Automatic Imaging Telescope (KAIT; Li
et al.\ 2000; Filippenko et al.\ 2001) at Lick Observatory. These data
and their implications are thoroughly discussed by \citet{leon01b}.

\subsection{SN~1998S}
\subsubsection{X-ray}
Our criterion for \chandra\ observations of SNe beyond 10~Mpc was the
detection of radio emission greater than 1~mJy at 6~cm.  SN~1998S met
this criterion on 1999 October 28 \citep{vandyk99}, and subsequent
reanalysis of early radio observations indicated very weak detection
at 8.46 GHz as early as 1999 January 07 \citep{lacey01}.  We observed
with \chandra\ on 2000 January 10, March 7, August 1, 2001 January 14,
and October 17 (days 678, 735, 882, 1048, and 1324 since optical
discovery).  Similar to the observations of SN~1999em, these data were
taken with the ACIS-S3 chip at nominal frame time (3.2 sec) in TE mode
and telemetered in F mode.  A summary of the observations is listed in
Table~\ref{tab:obs98s-xray}.

\subsubsection{Radio}
We observed SN~1998S over 20 times with the VLA at frequencies ranging
from 22.48 GHz to 1.46 GHz and ages from 89 to 1059 days after the
discovery date of 1998 March 03 \citep{lacey01}.  All observations yielded only upper
limits until 1999 January 07, age 310 days, when a weak detection was
obtained.  Since that time, monitoring has continued and SN~1998S has
increased in flux density at all mid-cm
wavelengths.

\subsubsection{Optical}
On 1998 March 7, just 5 days after the discovery of SN~1998S, optical
spectropolarimetry was obtained of the supernova with LRISp on the
Keck-II 10-m telescope \citep{leon01a}.  Total flux spectra were
obtained over the first 494 days after discovery, as follows: (a) at
Keck using LRIS on 1998 March 5, 6, and 27, and on 1999 January 6, and
(b) at the Lick 3-m Shane reflector using the Kast double spectrograph
on 1998 June 18, July 17 and 23, and November 25, and on 1999 January
10, March 12, and July 7.  See \citet{leon00} for details of the
optical observations and reductions.

\section{\chandra\ Data Reduction}
For each observation, we followed the data preparation threads
provided by the \chandra\ X-ray Center.  We used the \chandra\
Interactive Analysis of Observations (CIAO) software (version 2.2) to
perform the reductions, along with the CALDB~2.10 calibration files
(gain maps, quantum efficiency, quantum efficiency uniformity,
effective area).  Bad pixels were excluded, and intervals of bad
aspect were eliminated.

All data were searched for intervals of background flaring, in which
the count rate can increase by factors of up to 100 over the quiescent
rate.  The light curves of background regions were manually inspected
to determine intervals in which the quiescent rate could be reliably
calculated.  In accordance with the prescription given by the ACIS
Calibration Team, we identify flares as having a count rate $\ga$30\%
of the quiescent rate. Background flares were found in each of the
first four and the last observation of SN~1999em, lasting from
hundreds of seconds to over 15~ksec.  Only the second and fifth
observations of SN~1998S had such flares, lasting $\sim$300~sec and
$\sim$14~ksec, respectively. The event lists were filtered to exclude
the time intervals during which flares occurred.  Effective
observation times after filtering are listed in
Tables~\ref{tab:obs99em-xray} and \ref{tab:obs98s-xray}.

In addition, we have performed a reprocessing of all data so as not to
include the pixel randomization that is added during standard
processing.  This randomization has the effect of removing the
artificial substructure (Moir\'{e} pattern) that results as a
byproduct of spacecraft dither.  Since all of our observations
contained a substantial number of dither cycles (one dither cycle has
a period of $\sim$1000~sec), this substructure is effectively washed
out, and there is no need to blur the image with pixel randomization.
Removing this randomization slightly improves the point spread
function (PSF).

The source spectra were extracted with the CIAO tool {\it dmextract}.  We
excluded events with a pulse invariant (PI) of either 0 (underflow
bins) or 1024 (overflow bins).  The region of extraction was
determined from the CIAO tool {\it wavdetect}, a wavelet-based source
detection algorithm which can characterize source location and extent.
In each observation, the region determined by {\it wavdetect} for the
SN is consistent with that of a point source.  The data were
restricted to the energy range of 0.4--8.0~keV for the purposes of
spectral analysis since the effective area of ACIS falls off
considerably below 0.4~keV and the increased background above 8~keV
makes this data unreliable.

\begin{figure*}
\plotone{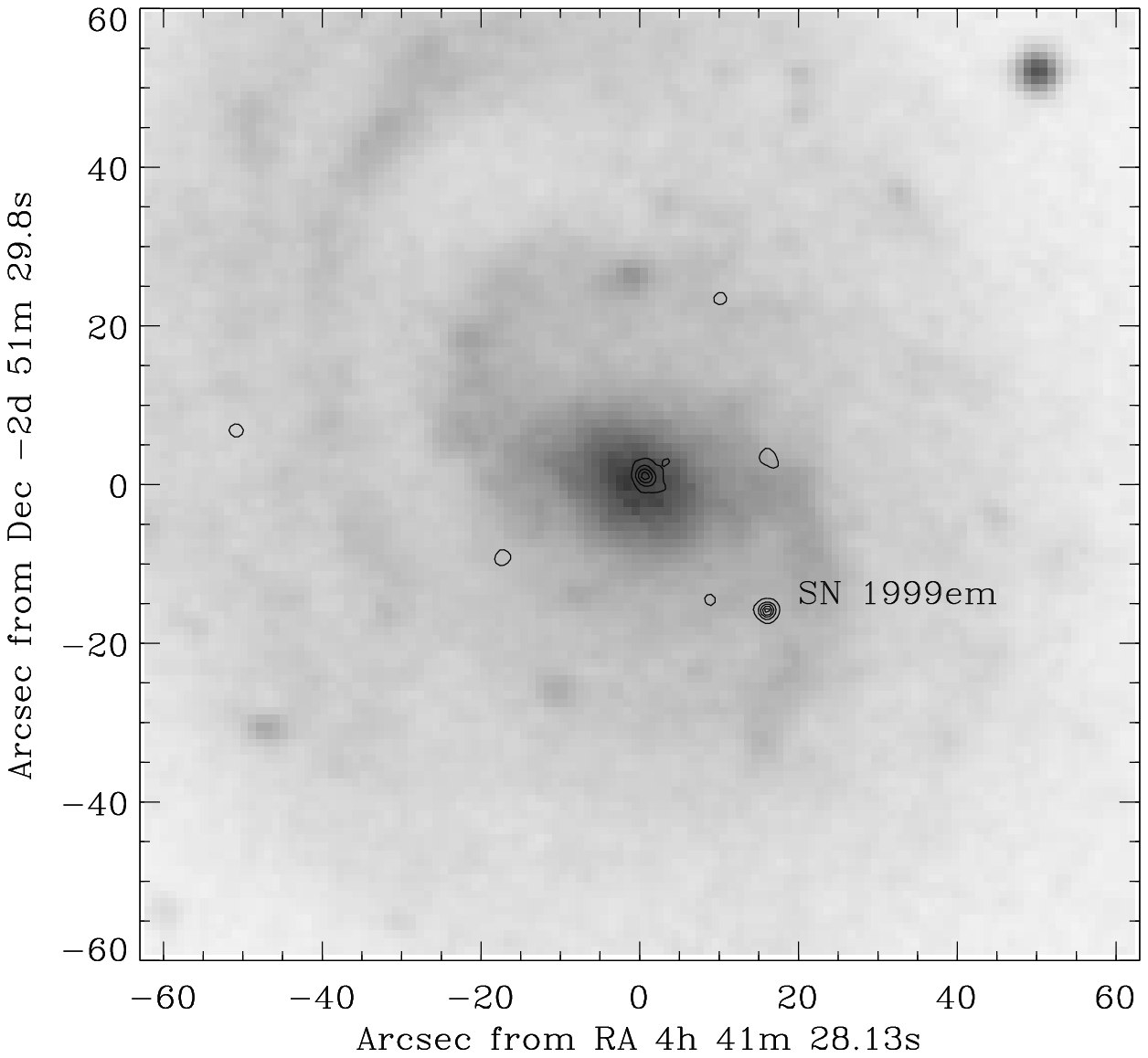}[th]
\caption{NGC~1637 in optical and X-rays.  X-ray contours from the
first \chandra\ observation of SN~1999em are overlaid on a DSS image
of the host galaxy. \label{fig:ngc1637}}
\end{figure*}

\section{Results}
\subsection{SN~1999em}
\subsubsection{X-ray}
The source was detected on all occasions, and an overlay of the X-ray
contours from the first observation on an optical image of the host
galaxy is shown in Fig.~\ref{fig:ngc1637}.  The {\it wavdetect}
position\footnote{This is the average position of observations 2--6,
in which the standard deviations were $\sigma_{\rm RA}=0\farcs03$ and
$\sigma_{\rm Dec}=0\farcs4$.  Observation~1 had a known aspect error
and was off by $\sim$3\arcsec.  This offset was corrected for
Fig.~\ref{fig:ngc1637}.}  (based solely on the \chandra\ aspect
solution) is 4\hr41\minute27\fs16, $-$2\degr51\arcmin45\farcs6, which is
within 1\arcsec\ (roughly the \chandra\ pointing accuracy) of the
radio and optical positions.  The low flux (\Fx\ $\la
10^{-14}$~\ergcms) resulted in few total counts (see Table~1).  We
have fit both the \mekal\ and thermal Bremsstrahlung models to our
data.  The \mekal\ model is a single-temperature, hot diffuse gas
\citep{mewe85} with elemental abundances set to the solar values of
\citet{andgrev89}.  We have calculated the column density
\citep{predschm95} to be $N_{\rm H}=6.1\times 10^{20}$~\pcmsq\ based
upon a value of $E_{(B-V)}=0.05$ \citep{baron00}.  For both models,
only the temperatures and overall normalizations were allowed to vary.
In \xspec\ \citep{arnaud96}, we performed maximum likelihood fits on
the unbinned data using Cash statistics \citep{cash79}.  These fits
should be insensitive to the number of counts per channel and are thus
appropriate for fitting low-count data. The best-fit observed
temperatures are shown in Table~\ref{tab:fits99em}.  As expected, the
uncertainties in fit parameters for such low-count data are large.

\begin{figure*}
\plotone{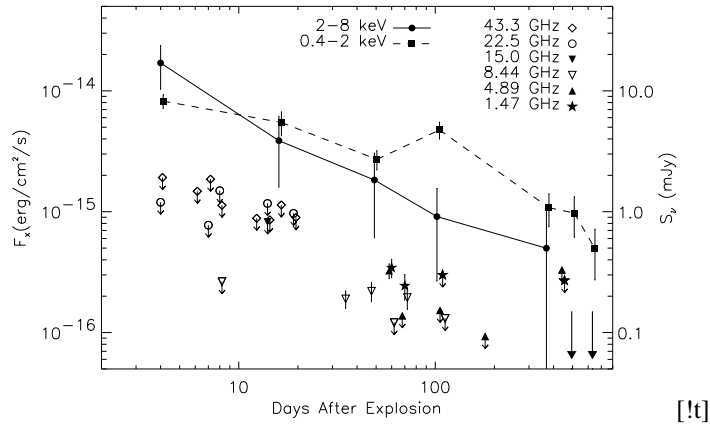}[!t]
\caption{X-ray and radio light curves of SN~1999em.  The detections
are shown with error bars while the upper limits are shown with
downward arrows.  Radio flux densities are plotted with the appropriate
symbols.  The absorbed X-ray flux is plotted in both the 2--8~keV
({\it solid line}) and 0.4--2~keV ({\it dashed line}) bands.  The hard X-rays
follow a steady decline, but the soft X-rays double from the third
observation to the fourth.} \label{fig:lc99em}
\end{figure*}

We have also used flux ratios to characterize the source.  The fluxes
were calculated by manually integrating the source spectra, i.e., they
are model-independent, absorbed fluxes.  The first observation had a
flux ratio of \Fx[2--8~keV]/\Fx[0.4--2~keV] $= 2.1 \pm 0.9$, but later
observations were softer (Table~\ref{tab:obs99em-xray}).  Somewhat
surprisingly, the total flux nearly doubled from the third observation
to the fourth, despite the continued decline of the high-energy
X-rays.  The multicolour X-ray lightcurve is shown in
Fig.~\ref{fig:lc99em}.

\subsubsection{Radio}
After 15 unsuccessful VLA observations of SN~1999em at ages 4 to 19~days
after the assumed explosion date of 1999 October 28,
the SN was finally detected on day 34 with 0.190 mJy flux density at
8.435~GHz.  Subsequent observing on days~46, 60, and 70 indicated that
SN~1999em remained detectable at a similar flux density, near the
sensitivity limit of the VLA, at 8.435, 4.885, and 1.465~GHz.  The
data are shown, together with the x-ray results, in
Fig.~\ref{fig:lc99em}. Since day~70, no radio detection of SN~1999em
in any cm wavelength band has been obtained.  SN~1999em is among only
a few radio detected Type II-P, and its behavior is difficult to
compare with other radio observations of the same type SN.  An
estimated spectral luminosity of SN~1999em at 6 cm peak on day
$\sim$34 of $L_{\rm 6cm~peak} \sim 2.2 \times 10^{25}$ erg s$^{-1}$
Hz$^{-1}$ makes it the least radio luminous RSN known except for the
peculiar, and very nearby, SN~1987A.
The radio position of the emission from
SN~1999em is 4\hr41\minute27\fs157, $-$2\degr51\arcmin45\farcs83
(J2000).  The details of the observations and analysis will be found
in \citet{lacey01}.

\subsubsection{Optical}
The spectropolarimetry of SN~1999em provided a rare opportunity to
study the geometry of the electron-scattering atmosphere of a
``normal'' core-collapse event at multiple epochs \citep{leon01a}. A very low but
temporally increasing polarization level suggests a substantially
spherical geometry at early times that becomes more aspherical at late
times as ever-deeper layers of the ejecta are revealed.  When modeled
in terms of oblate, electron-scattering atmospheres, the observed
polarization implies an asphericity of at least 7\% during the period
studied. We speculate that the thick hydrogen envelope intact at the
time of explosion in SNe II-P might serve to dampen the effects of an
intrinsically aspherical explosion. The increase in asphericity seen
at later times is consistent with a trend recently identified
\citep{wang01} among stripped-envelope core-collapse SNe: the deeper
we peer, the more evidence we find for asphericity.  The natural
conclusion that it is an {\it explosion} asymmetry that is responsible
for the polarization has fueled the idea that some core-collapse SNe
produce gamma-ray bursts (GRB; e.g., \citealt{bloom99}) through the
action of a jet of material aimed fortuitously at the observer, the
result of a ``bipolar,'' jet-induced, SN explosion
\citep{kho99,whee00,maeda00}.  Additional multi-epoch
spectropolarimetry of SNe II-P are clearly needed, however, to
determine if the temporal polarization increase seen in SN~1999em is
generic to this class.

\subsection{SN~1998S}
\subsubsection{X-ray}
\begin{figure*}
\plotone{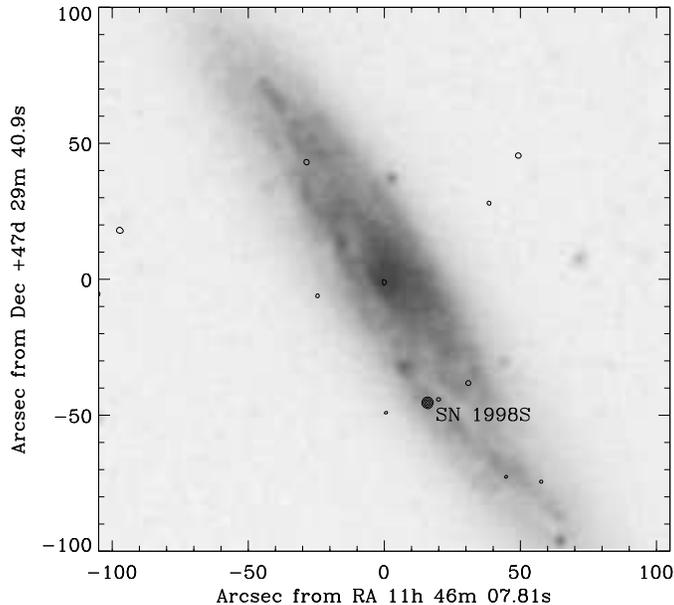}
\caption{NGC~3877 in optical and X-rays.  X-ray contours from the
first \chandra\ observation of SN~1998S are overlaid on a DSS image
of the host galaxy. \label{fig:ngc3877}}
\end{figure*}

\begin{figure*}[t!]
\plotone{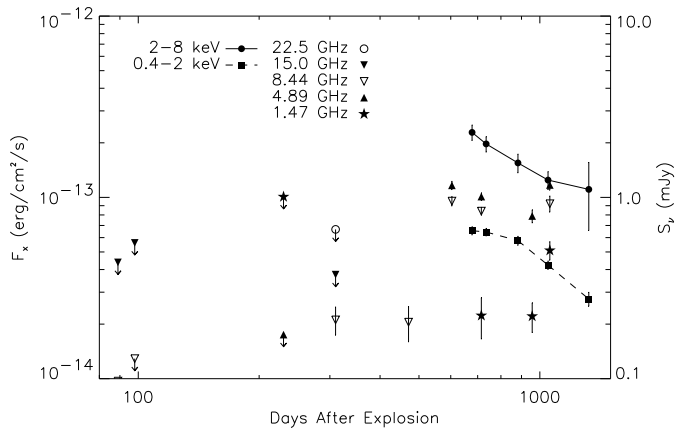}
\caption{X-ray and radio light curves of SN~1998S. The detections are
shown with error bars while the upper limits are shown with downward
arrows.  The absorbed X-ray flux declines in both the hard ({\it solid
lines}) and soft ({\it dashed lines}) bands.} \label{fig:lc98s}
\end{figure*}

The source was detected on all occasions, and an overlay of the X-ray
contours from the first observation on an optical image of the host
galaxy is shown in Fig.~\ref{fig:ngc3877}.  The {\it wavdetect}
position\footnote{This is the average position of all observations.
The standard deviations about the means were $\sigma_{\rm
RA}=0\farcs03$ and $\sigma_{\rm Dec}=0\farcs17$.}  (based solely on
the \chandra\ aspect solution) is 11\hr46\minute06\fs14,
47\degr28\arcmin55\farcs1 (J2000), which is within 0\farcs5 of the
radio and optical positions.  In addition to constructing multicolour
X-ray lightcurves (Fig.~\ref{fig:lc98s}), the high luminosity of
SN~1998S ($\Lx \approx 10^{40}$~\ergsec) allowed for basic spectral
fitting to be done for each of the first four observations; in the
fifth, there were too few counts for a similar type of analysis.  We
performed this in \xspec\ with the \vmek\ model, which is identical to
the \mekal\ model but allows individual elemental abundances to vary.
Using the reddening
$E_{(B-V)}=0.23$ \citep{leon00}, we calculated the hydrogen column
density to be $N_{\rm H}=1.36\times 10^{21}\ \pcmsq$
\citep{predschm95}.  We used a redshift of $z=0.003$
\citep{willick97}.

Although there were sufficient counts ($>$500 per observation) in each
of the first four observations for spectral modeling, there were few
counts per energy channel.  To mitigate the problems associated with
fitting such data, we grouped the data to contain a minimum number of
counts per bin.  Because there were virtually no counts above $\sim$
7~keV in any of the observations, we would ``smear'' the Fe line at
6.7~keV if we required more than $\sim$ 7 counts per bin.  Therefore,
we set the minimum number of counts per bin to the highest possible
value which would still retain the integrity of the Fe line.  This
turned out to be 5 counts per bin for the first observation, 7 for the
second, 5 for the third, and 7 for the fourth.

To reduce the number of free parameters in our fits, we froze the
abundance of He to its solar value.  We also linked the
abundances of N, Na, Al, Ar, and Ca to vary with C.  The remaining
elemental abundances (O, Ne, Mg, Si, S, Fe, and Ni) were allowed to
vary independently.  We first allowed the column density to vary in
our fits.  Since the best-fit results were all consistent with the
$N_{\rm H}$ obtained from the optical reddening, we fixed this
parameter at the calculated value.

\begin{figure*}
\resizebox{3in}{!}{\rotatebox{270}{\plotone{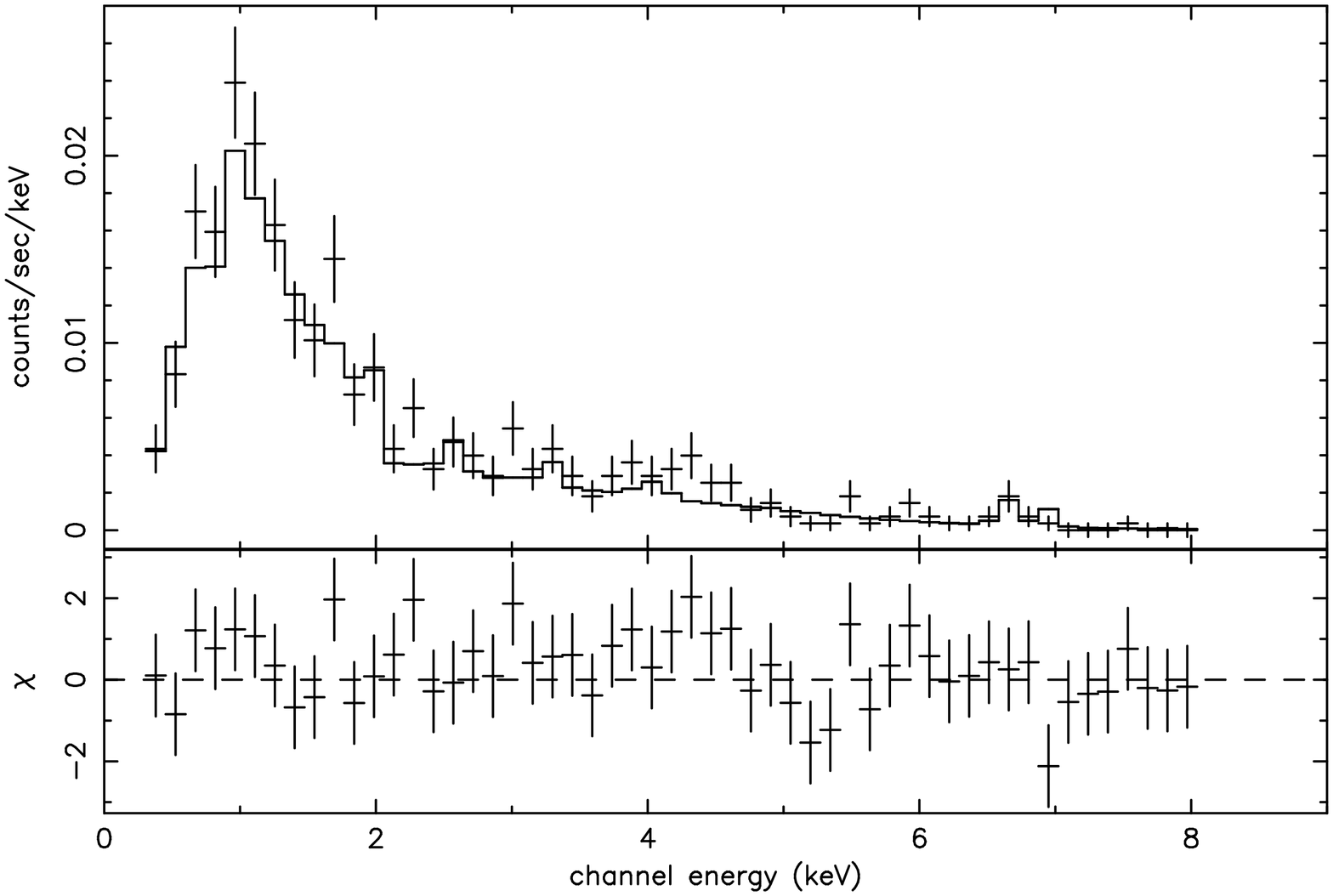}}}
\hspace{0.2in}
\resizebox{3in}{!}{\rotatebox{270}{\plotone{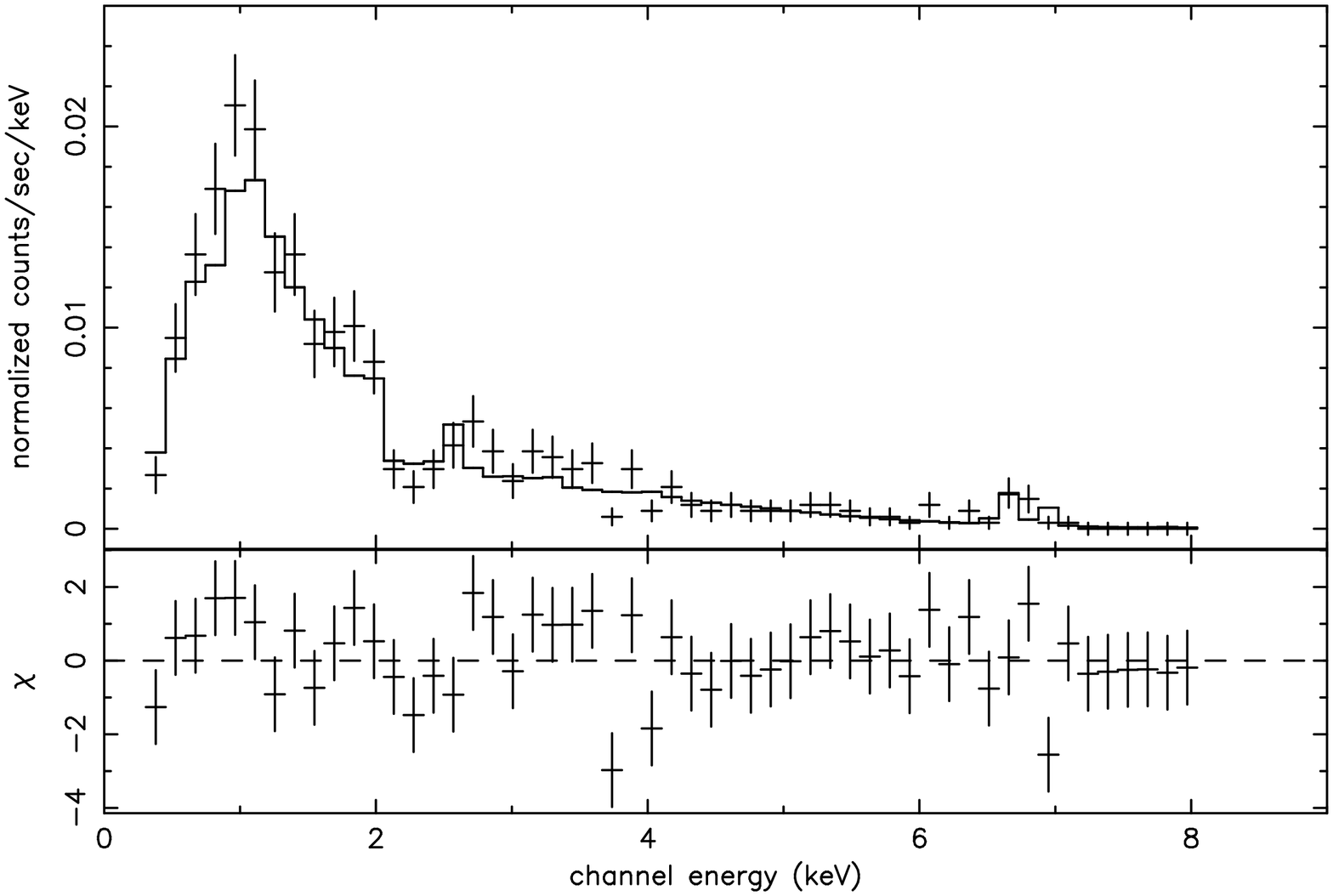}}}

\vspace{0.2in}
\resizebox{3in}{!}{\rotatebox{270}{\plotone{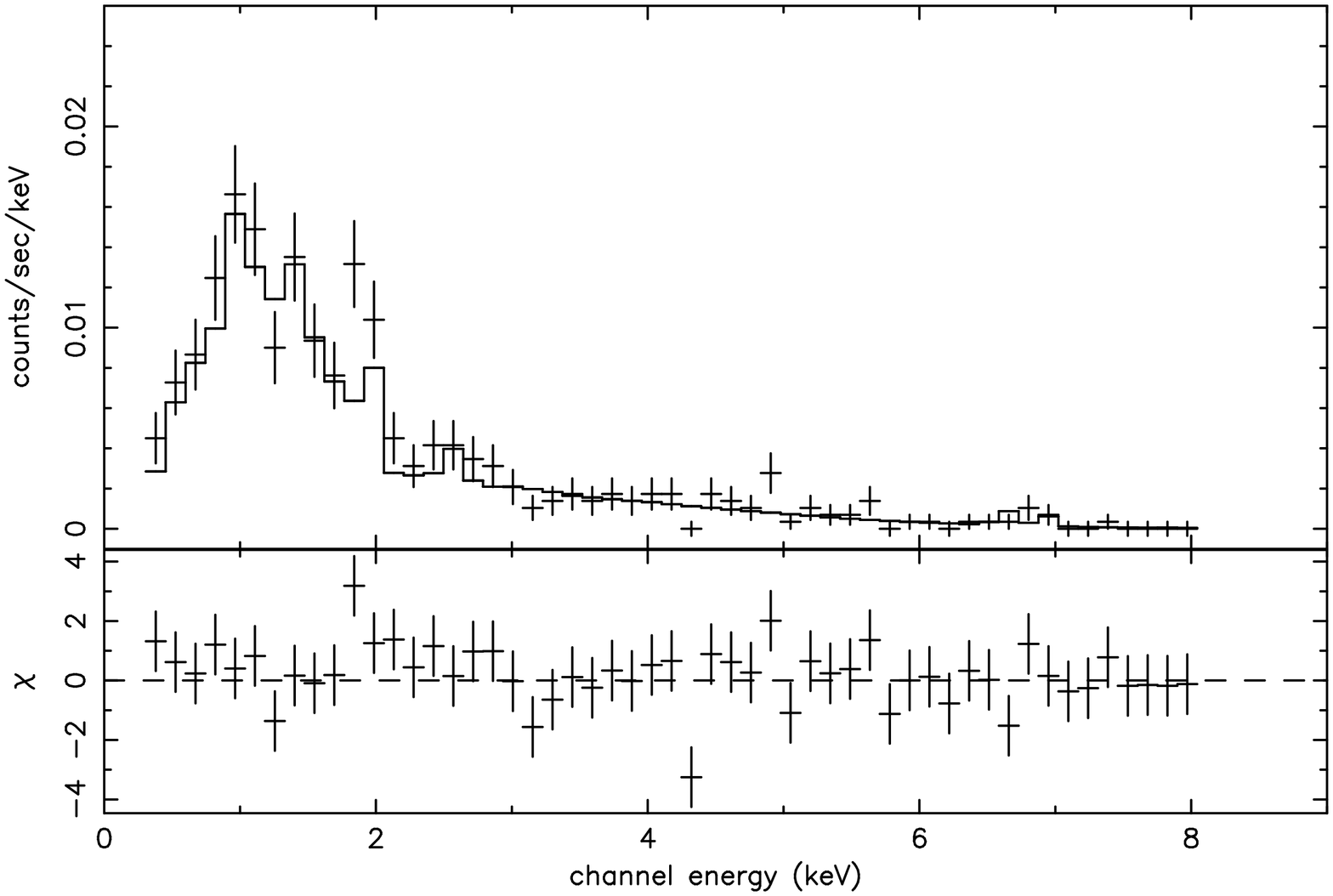}}}
\hspace{0.2in}
\resizebox{3in}{!}{\rotatebox{270}{\plotone{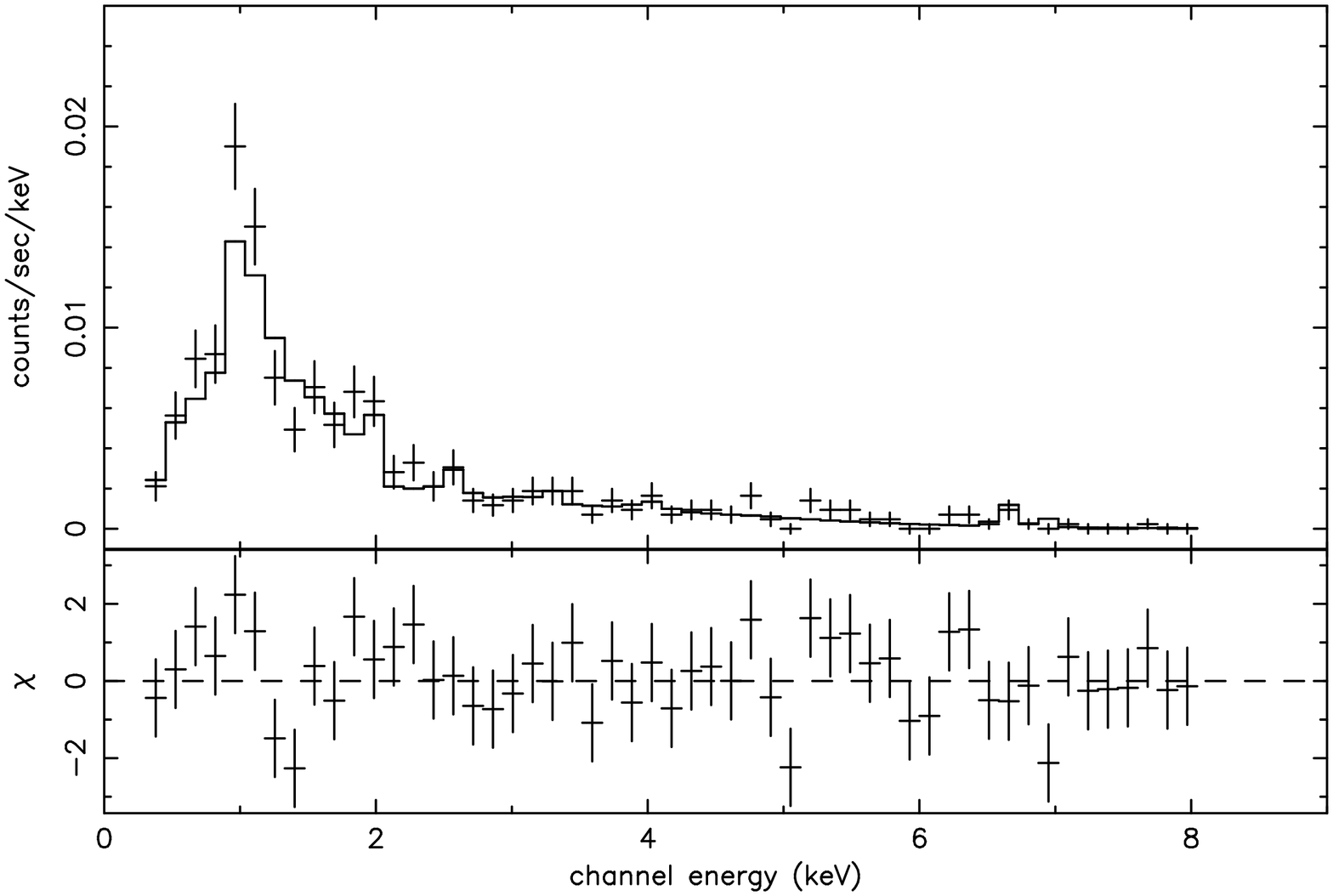}}}
\caption{Spectra of SN~1998S.  The first observation is in
the upper left, the second in the lower left, the third in the upper
right, and the fourth in the lower right.  Best-fit \vmek\ models are
overplotted on evenly binned data.}\label{fig:spec98s}
\end{figure*}

We then performed $\chi^2$ fitting using Gehrels weighting for the
statistical error \citep{gehrels86}, which is appropriate for low
count data.  We also performed maximum likelihood fits to the unbinned
data using Cash statistics \citep{cash79}.  The maximum likelihood
fits were in good agreement with the $\chi^2$ fits, and we report the
$\chi^2$ results in Table~\ref{tab:fits98s}.  We note, however, that
fitting with so few counts per bin may give misleading $\chi^2$
values.  For ease of viewing, we have plotted the best fit models
against evenly binned data (Fig.~\ref{fig:spec98s}).

In order to bring out the line features, we have summed the individual
spectra.  We feel justified in doing this because the best-fit model
parameters from the four individual observations with enough counts
for individual modeling are consistent with each other, indicating
that the spectral shape is not changing much, only the overall
normalization (luminosity).  We grouped the summed spectrum to
$\geq$15 counts per bin for spectral fitting.  Grouping to more counts
per bin than that would have washed out some of the weaker line
features.  The individual response functions for each observations
were combined in a weighted average based on exposure times.

\begin{figure*}[t]
\resizebox{\textwidth}{!}{\rotatebox{270}{\plotone{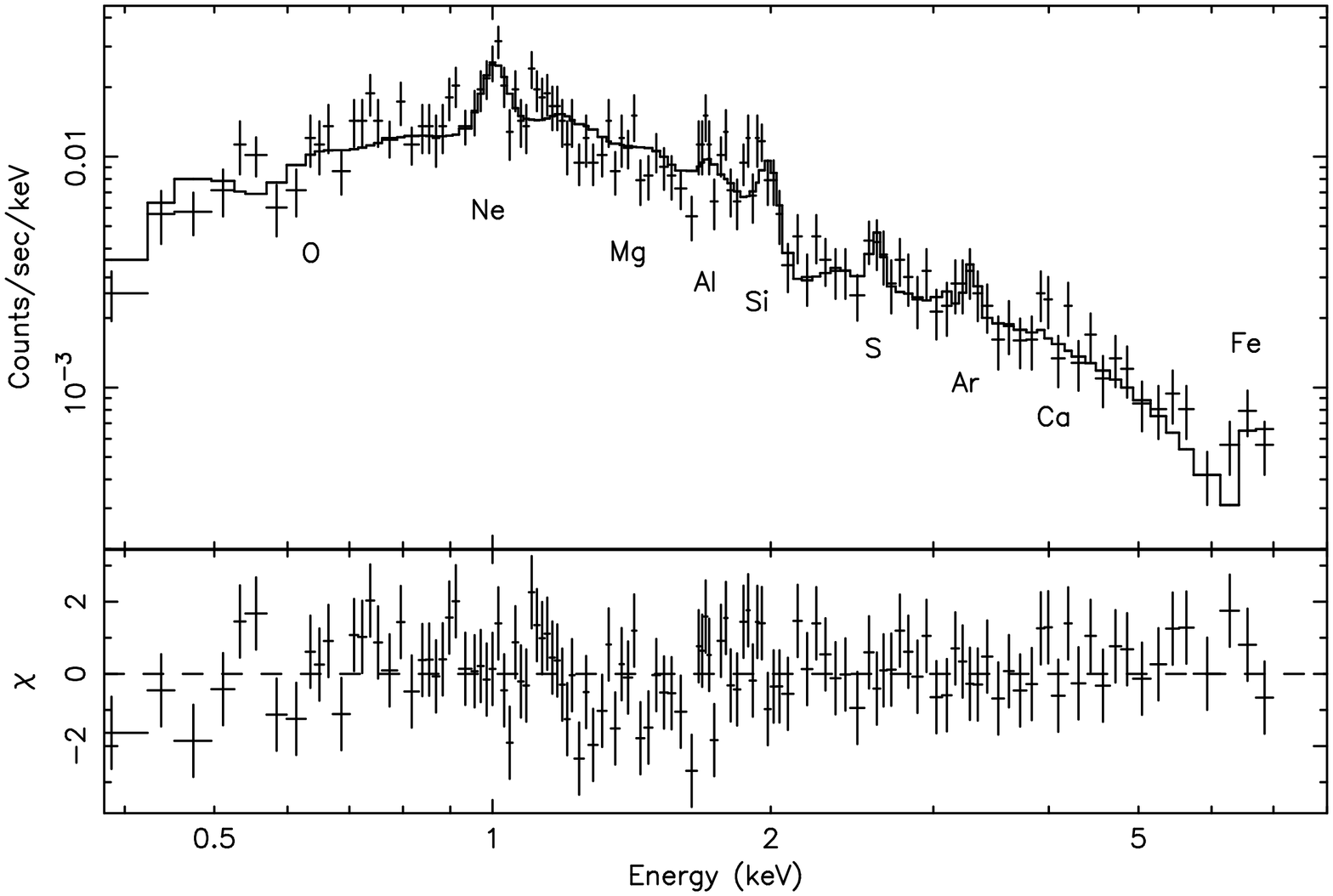}}}
\caption{\vmek\ fit to the summed spectrum of SN~1998S.  Labels
indicate approximately where emission lines for each element would be
observed.}
\label{fig:summed98s}
\end{figure*}

We fit four models to this summed spectrum: a single \mekal\ plasma
with solar abundances (1-T \mekal); two \mekal\ plasmas with solar
abundances (2-T \mekal); a single \mekal\ plasma with variable
abundances (all elements allowed to vary except He; 1-T \vmek); and a
single \mekal\ with solar abundances plus a single \mekal\ with
variable abundances (2-T \vmek).  The 1-T \mekal\ model did not fit
the data well, with a $\chi^2$ of 181 for 124 degrees of freedom
(dof).  Tests of the other three model fits against this one using the
F-statistic indicate that only the 1-T \vmek\ fit was a significant
improvement (required at the 3.8~$\sigma$ level).  Most of the
abundances in this model are not well constrained (especially those of
C, N, Na, and Ni).  The uncertainty in each abundance was determined
in the following way.  The abundance was changed from the best-fit
value, the model was refit (with all other parameters allowed to
vary), and the resultant $\chi^2$ was compared to the best-fit
$\chi^2$.  This was repeated until the difference in $\chi^2$ was
2.706, which gave the 90\% confidence interval for the given
parameter.  Table~\ref{tab:summed98s} lists the results.  The best-fit
temperature was $kT = 9.8^{+2.2}_{-1.8}$~keV, and the fit statistic
was $\chi^2 = 139$ for 116 degrees of freedom.
Fig.~\ref{fig:summed98s} shows the summed spectrum and best-fit \vmek\
model with labels indicating the rough positions of elemental lines.

\subsubsection{Radio}
Since its initial radio detection at 8.46~GHz on 1999 January 07 (310 days
after optical discovery), SN~1998S has increased at all mid-cm
wavelengths.  Modeling indicates that it should reach its 6 cm flux
density peak at age $\sim$1000 days, at roughly the epoch of the most
recent observations on 2001 January 25.  Assuming that is the case,
SN~1998S is behaving like other radio-detected Type IIn (SNe~1986J and
1988Z both peaked at 6 cm wavelength at age $\sim$1000 days)
although it will be the least radio luminous at 6 cm peak ($L_{\rm
6cm~peak} \sim 3.6 \times 10^{26}$ erg s$^{-1}$ Hz$^{-1}$) of the six
Type IIn RSNe known (SNe~1978K, 1986J, 1988Z, 1995N,
1997eg, and 1998S).  The radio data are shown together with the
X-ray results in Fig.~\ref{fig:lc98s}.

Under the Chevalier model of a dense CSM established by a slow
pre-supernova stellar wind ($v_w$ = 10 km s$^{-1}$, $v_{\rm shock} =
10^4$ km s$^{-1}$, $T = 2 \times 10^4$ K; see, e.g.,
\citealt{weiler86}), such a radio luminosity is interpreted as
indicative of a mass-loss rate of $\sim 2 \times 10^{-4}$~\ml. The
radio position of the emission from SN~1998S is 11\hr46\minute6\fs140,
47\degr28\arcmin55\farcs45.  The details of the observations and
analysis will be found in \citet{lacey01}.  It should be noted that
the best fit to the radio data requires significant clumping or
filamentation in the CSM, as was also found for the Type IIn SNe 1986J
and 1988Z \citep{weiler90,vandyk93}.

\subsubsection{Optical}
SN~1998S exhibits a high degree of linear polarization,
implying significant asphericity for its continuum-scattering
environment \citep{leon00}.  Prior to removal of the interstellar
polarization, the polarization spectrum is characterized by a flat
continuum (at $p \approx 2\%$) with distinct changes in polarization
associated with both the broad (symmetric, half width near zero
intensity $\gtrsim 10,000$ km s$^{-1}$) and narrow (unresolved, full
width at half maximum $< 300$ km s$^{-1}$) line emission seen in the
total flux spectrum.  When analyzed in terms of a polarized continuum
with unpolarized broad-line recombination emission, an intrinsic
continuum polarization of $p \approx 3\%$ results, suggesting a global
asphericity of $\gtrsim 45\%$ from the oblate, electron-scattering
dominated models of \citet{hof91}.  The smooth, blue continuum evident
at early times is inconsistent with a reddened, single-temperature
blackbody, instead having a color temperature that increases with
decreasing wavelength.  Broad emission-line profiles with distinct
blue and red peaks are seen in the total flux spectra at later times,
suggesting a disk-like or ring-like morphology for the dense ($n_e
\approx 10^7 {\rm\ cm^{-3}}$) circumstellar medium, generically
similar to what is seen directly in SN~1987A, although much denser and
closer to the progenitor in SN~1998S.  Such a disk/ring-like
circumstellar medium may have formed from a merging of a
binary-companion star that ejects the common-envelope material in the
direction of the orbital plane of the binary system
\citep{nomoto95}.

\section{Discussion}
\subsection{SN~1999em}
To interpret the X-ray and radio data on SN 1999em, we use
the circumstellar interaction model proposed by \citet{chev82},
and elaborated by \citet{chevfran94} and \citet{flc96}.
The outer supernova ejecta are taken to be freely expanding, with
a power law density profile
\begin{equation}
\rho_{sn}=At^{-3}(r/t)^{-n},
\end{equation}
where $A$ and $n$ are constants.  SN 1999em was a Type II-P SN,
so we expect that the star exploded as a red supergiant with most of
its hydrogen envelope.  The value of $n$ for such a star is usually
taken to be in the range $7-12$ \citep{chev82}, although
\citet{matz99} recently found that the explosion of a red supergiant
leads to $n=11.9$ at high velocity.  In the case of SN~1999em, we have
some additional information from modeling of the optical spectrum.
\citet{baron00} were able to model spectra on 1999 October 29 and
November 4/5 with $n=7$, 8, or 9.  The modeling of the data for the
earlier time suggests a relatively flat density profile ($n=7$).  The
spectra on 1999 October 29 also show evidence for a secondary absorption
feature at 20,000~\kmsec, implying the presence of denser material in
the cool ejecta.  One possible origin for high velocity, dense ejecta
is the shell that can form as a result of the diffusion of radiation
at the time of shock breakout (Chevalier 1976); if this is the origin,
it represents the highest velocity material in the ejecta.  We assume
that the outer density profile can be approximated as a power law out
to this high velocity and take $n=7$ or $9$.  The value of $A$ is
equivalent to specifying the density in the ejecta at some particular
velocity and age.  The model by \citet{baron00} for 1999 October 29 has
$\rho_{sn} =0.4\times 10^{-20}$~\gcm\ at 10,000~\kmsec\ at an age of 1
year.  This value is in approximate accord with the density found in
models with a broken power law density profile, mass of 10~\Msun,
and energy of $1\times 10^{51}$~ergs \citep{chevfran94}.

The density in the wind is $\rho_w=\dot M/(4\pi r^2 v_w)$, where $\dot
M$ is the mass-loss rate and $v_w$ is the wind velocity.  We take
$\dot M_{-6}=\dot M/10^{-6}$~\ml\ and $v_{w1}=v_w/10$~\kmsec\ as
reference values.  The interaction of the supernova with a freely
expanding wind leads to a shocked shell with radius $R\propto t^{0.8}
(n=7)$ or $t^{6/7} (n=9)$.  This shell is the source of the X-ray
emission.  The low observed X-ray luminosity suggests that the shocked
gas is not radiative; this can be checked for consistency after a
model is produced.  Under these conditions, the X-ray luminosity is
expected to be dominated by emission from the reverse shock region,
which is relatively cool, as observed.  The luminosity of the shell
can be estimated from eq.~(3.10) of FLC96
\begin{equation}
L_{rev}=2.0\times 10^{35}\zeta (n-3)(n-4)^2 T_8^{-0.24} e^{-0.116/T_8}
\left(\dot M_{-6}\over v_{w1}\right)^2 V_4^{-1}
\left(t\over 10{\rm~days}\right)^{-1} {\rm~ ergs~s^{-1}~keV^{-1}},
\end{equation}
where $\zeta$ is 0.86 for solar abundances and 0.60 for $n({\rm
He})/n({\rm H})=1$, $T_8=T/10^8$~K, and $V_4$ is the peak velocity in
units of $10^4$~\kmsec.  \citet{baron00} find possible evidence for
enhanced He in SN 1999em, but this is uncertain and is probably not
expected in a Type II-P SN.  We take $\zeta= 0.86$; the
uncertainty is small.  On 1999 November 1, which we take to be day~4, the
observed luminosity (0.4--8~keV) is $2\times 10^{38}$~\ergsec\ and the temperature
is 5.0~keV.  With $V_4=1.5$, we find that $\dot M_{-6}/ v_{w1}\approx
1$ $(n=9)$ or 2 $(n=7)$, quite similar to the value of $\dot
M_{-6}/ v_{w1} \sim 2$ obtained above from the radio observations.
With this mass-loss rate, the cooling time for the gas, as deduced
from eq. (3.7) of FLC96, is $31$~days for $n=9$ and is longer for the
$n=7$ case, which justifies our use of the adiabatic expression for
the X-ray luminosity.  The cooling time grows more rapidly than the
age, so we expect non-radiative evolution through the time of our
observations.  If cooling were important, a dense shell would form
that could absorb X-ray emission from the reverse shock wave, as
appeared to occur in the initial evolution of SN~1993J (FLC96; Suzuki
\& Nomoto 1995).  Reradiation of the absorbed emission can give broad
optical emission lines, as observed in SN~1993J at ages $\ga 1$~year.
Such lines were not observed in SN~1999em, which is consistent with
non-radiative evolution.

From eq.~(2.2) of FLC96 with $n=9$, we find that the maximum velocity
for $n=9$ is
\begin{equation}
V=\left(4\pi \rho_o t_o^3 v_o^9 v_w\over 15 \dot M\right)^{1/7} t^{-1/7},
\end{equation}
where $\rho_o$ specifies an ejecta density at a particular time $t_o$
and velocity $v_o$, as described above.  At $t=4$ days with $\rho_o=
0.4\times 10^{-20}$~\gcm\ at $v_o=10,000$~\kmsec\ and $t_o=1$~year and
$\dot M_{-6}/ v_{w1}= 1$, we have $V=13,000$~\kmsec.  For the $n=7$
case with $\dot M_{-6}/ v_{w1}= 2$, we have $V=14,000$~\kmsec\ and
$V\propto t^{-0.2}$.  These velocities are lower than the highest
velocities deduced by \citet{baron00} on 1999 October 29 (day 1). However,
there is some uncertainty in the high velocity, and there may have been
rapid evolution at early times.

The shock velocity determines the temperature of the reverse shock
region.  From eqs.~(3.1) and (3.2) of FLC96, we have $T_{\rm rev}=2.4
V_4^2$~keV for $n=9$, so that with $V_4=1.3$, $T_{\rm rev}=4.1$~keV.
For $n=7$, we have $T_{\rm rev}=4.8 V_4^2$~keV so $T_{\rm rev}=9.4$~keV for $V_4=1.4$.  The observed temperature on 1999 November 1 falls
between these two cases.  The predicted evolution of the temperature
is $T\propto V^2\propto t^{-2/(n-2)} \propto t^{-2/7}$ $(n=9)$ and
$\propto t^{-0.4}$ $(n=7)$.  Table~\ref{tab:fits99em} compares this
predicted evolution with that observed.  The observations show a clear
cooling, as expected. An additional factor could be the lack of
electron-ion equilibration at the reverse shock front; if it is not
achieved, the electrons are cooler than the above estimate.  From
eq.~(3.6) of FLC96, the expected conditions are such that the
equilibration timescale is comparable to the age throughout the
evolution, so that no firm conclusion can be drawn.  The fact that
cooling of the emission is clearly observed argues for a relatively
flat density profile for the ejecta, as also found by \citet{baron00}
from optical spectroscopy.

An estimate of the expected evolution of the total X-ray luminosity,
\Lx, is given by \citet{chevfran94}.  When free-free emission
dominates ($T\ge 4\times 10^7$~K), $\Lx\propto t^{-1}$, and when lines
dominate ($10^5 \le T < 4\times 10^7$~K), $\Lx\propto t^{-0.69}$
$(n=9)$ or $\propto t^{-0.56}$ $(n=7)$.  The observed evolution is
less steep than $t^{-1}$ and is closer to the expectation for line
emission.  The observed temperatures indicate that a transition occurred.
Again, the observations are in
approximate accord with expectations.

One aspect of the X-ray observations that is not compatible with
the simplest wind interaction model is the increase of the soft X-ray
flux on 2000 February 2 (day~102, see Fig.~\ref{fig:lc99em}).  In our
model, the emission is primarily from the reverse shocked ejecta, so the
increase could be due to a region of increased density in the ejecta.
At this age, the velocity of the ejecta entering the reverse shock
front is $\sim 8,000$~\kmsec.  Alternatively, the emission could be
from an inhomogeneity in the circumstellar wind.  The smooth wind is
shocked to a temperature $\sim 10^9$~K, so that a density contrast of
$\sim 100$ is needed for the inhomogeneity.  The inhomogeneity must be
large to affect the total luminosity, but a substantial change in
stellar mass-loss characteristics at only $10^{16}$~cm from the star
is unlikely.  The fact that the change is transitory also argues
against a change in mass-loss properties at this point. 

\begin{figure*}
\plotone{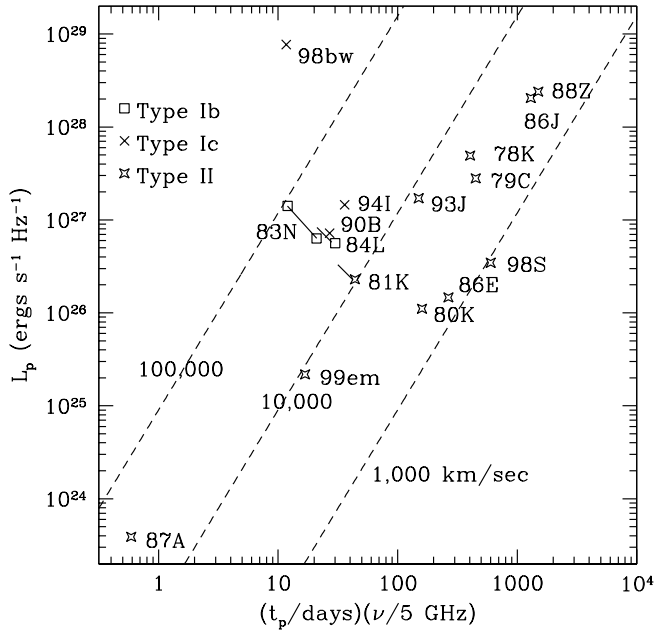}
\caption{The peak spectral radio luminosity vs the product of the time
of the peak and the frequency of the measurement, based on Fig. 4 of
Chevalier (1998) with the addition of SN 1998bw, SN 1998S, and SN
1999em.  The observed supernovae are designated by the last two digits
of the year and the letter, and are of Types II ({\it stars}), Ib
({\it squares}), and Ic ({\it crosses}).  The dashed lines show the
mean velocity of the radio shell if synchrotron self-absorption is
responsible for determining the flux density peak; an electron
spectrum $N(E)\propto E^{-2.5}$ is assumed.}
\label{fig:rad0307}
\end{figure*}

Although SN~1999em was detected as a weak radio source, the radio
light curves are not well defined because of the very low radio flux
density.  The early observations were not of sufficient sensitivity to
detect the source so the time of light curve peak and the early
absorption process are also not well defined.  The 8.44~GHz
observations are the most useful in this regard.  At late times, the
observed flux evolution of other radio supernovae is $\propto
t^{-0.7}$, or steeper \citep{weiler86}, so that the 8.44~GHz upper
limit on day~7 probably indicates that absorption was significant at
that time.  With this value for the time of peak flux at 8.44~GHz
(actually a lower limit), we can place SN~1999em on a diagram of peak
radio luminosity vs. time of peak (Fig.~\ref{fig:rad0307}).  The
diagonal lines show the velocity of the shocked shell if synchrotron
self-absorption were the important process for the early absorption
\citep{chev98}.  The position of SN~1999em gives about
$10,000$~\kmsec, or less, as the shell velocity.  Thus, it is just
possible that synchrotron self-absorption was important at early times
but more likely that another process, probably free-free absorption by
the unshocked wind, was dominant.  From eq.~(2.3) of FLC96, the
8.44~GHz free-free optical depth on day~7 with $V_4$ and $\dot M_{-6}/
v_{w1}$ as deduced above is $\tau=0.2 T_5^{-3/2}$ $(n=9)$ or $0.5
T_5^{-3/2}$ $(n=7)$, where $T_5$ is the temperature of the unshocked
wind in units of $10^5$~K.  The $n=7$ case is somewhat more consistent
with the observations because of the higher mass-loss rate deduced for
this case.

The free-free absorption of the radio emission depends on the
circumstellar temperature.  \citet{eastman94} find from a
simulation of a Type II-P explosion an effective temperature at
outbreak of $T_{\rm eff}= 1.8\times 10^5$~K and luminosity $9\times
10^{44}$~\ergsec. This temperature is similar to that used in one model
by \citet{lf91} for SN~1987A (their Table 1). Their
model with $\dot M_{-6}/v_{w1} = 0.4$ gives a wind temperature of
$1.2\times 10^5$~K for $T_{\rm eff} = 2\times 10^5$~K. Because of the low
wind density, this temperature is not sensitive to the mass-loss rate
and is mainly set by the outburst radiation temperature.  An
approximate temperature in the wind of SN~1999em should therefore be
$T_5\approx 1.2$. The temperature at $\sim$30~days should not be very
different because of the low wind density. X-ray heating by the shock
has not been included here, but is probably not important because of
the low X-ray luminosity.


Overall, the X-ray and radio observations of SN~1999em are consistent
with the non-radiative interaction of supernova ejecta with a power law
profile ($n \sim 7$) interacting with a pre-supernova wind with $\dot
M_{-6}/ v_{w1}\approx 2$.  We estimate the accuracy of this mass-loss
density to be a factor of 2.  The X-ray emission is primarily from the
reversed shocked ejecta, and the model can account for the luminosity,
temperature, and temperature evolution of the emission.  It does not
account for the large soft X-ray flux observed on day~102.  The radio
observations are compatible with this general picture and suggest a
comparable mass-loss rate.

\subsection{SN 1998S}
The high X-ray luminosity of SN~1998S places it in the same class as
the X-ray bright Type IIn SNe~1978K, 1986J, 1988Z, and 1995N.
\citet{fox00} discussed X-ray observations of SN~1995N and summarized
the results on the other SNe.  SN~1995N had an X-ray luminosity of
$1.5\times 10^{41}$~\ergsec\ at an age of 2.0~years.  SN~1998S is at
the low end of the X-ray luminosities spanned by these sources

The optical emission from SN 1998S shows both narrow and broad lines
that can be attributed to circumstellar interaction
\citep{fas00,leon00,chug01}.  The narrow lines are presumably from dense
circumstellar gas that is heated and ionized by the supernova
radiation.  Although high velocities ($V_4\approx 1$) are observed in
the early evolution of SN~1998S, by day~494 the maximum observed
velocities were $V_4 = 0.5-0.6$ \citep{leon00}.  These velocities are
much higher than the estimated velocity of gas at the photosphere,
indicating that the emission is from circumstellar interaction.

The interpretation of the X-ray emission from Type IIn supernovae is
not straightforward because of the possibility that the emission is
from shocked circumstellar clumps (see, e.g., \citealt{weiler90} and
\citealt{chug93} on SN~1986J and \citealt{vandyk93} on SN~1988Z). The
radio results for SN 1998S indicate that the CSM is likely clumpy or
filamentary \citep{lacey01}.  However, if our finding of heavy element
overabundances is correct, the X-ray emission is expected to be from
shocked SN ejecta.  If the interaction can be approximately
described by smooth ejecta running into a smooth circumstellar wind,
the temperature behind the reverse shock front is $T_{\rm rev}=2.4
V_4^2$~keV ($n=9$), $5.4 V_4^2$~keV ($n=7$), or $9.6 V_4^2$~keV
($n=6$).  It is difficult to produce the high observed temperature
($T\approx 10$~keV) unless there is unseen high velocity gas and the
ejecta density profile is relatively flat.  The observations do not
show the cooling that is expected as the shock interaction region
decelerates, but the range in time is not large enough for the effect
to be clearly visible.

The X-ray luminosity shows a clear decline with time, approximately
$\propto t^{-1.3}$.  This is in reasonable agreement with what is expected for
free-free emission from a non-radiative shock region ($t^{-1}$
dependence as discussed above), so we can again use eq.~(2) to
estimate the mass-loss rate.  The result is $\dot M_{-4}/ v_{w1} =
(1-2)$, where $\dot M_{-4}=\dot M/10^{-4}$~\ml, if $n=6$ and $V_4 =1$
(these estimates are implied by the high observed temperature of
$\sim$10~keV) on day~665, again comparable to that determined from
radio modeling of $\dot M_{-4}/ v_{w1} \sim 2$.  A high mass-loss rate
is required to produce the X-ray luminosity, but the cooling time is
still longer than the age.  The assumption of a non-radiative reverse
shock region is justified.  In this model, the X-ray luminosity of the
forward shock region is comparable to that from the reverse shock.
However, the higher temperature of the forward shock region implies
that the reverse shock dominates the emission in the energy band that
we observed.

The radio luminosity of SN 1998S is low for a RSN with a
late turn-on time (see Fig.~\ref{fig:rad0307}).  This implies that
synchrotron self-absorption was not important for the early
absorption.
This accounts for the basic features of the X-ray emission.  The flat
supernova density profile implies that much of the supernova energy is
in high velocity ejecta.  A similar deduction was made by
\citet{chug94} for SN~1988Z.  The heavy element overabundances require
that these elements be mixed out to a high velocity in the ejecta,
which has been shown to occur in Cas~A \citep{hughes00}.  In
theoretical models of aspherical explosions, heavy-element-rich
matter is naturally ejected at high velocities, which well explains
the peculiar late spectral features of SN~1998bw \citep{maeda00}.  The
polarization and spectral line profiles observed in SN~1998S
\citep{leon00} show that the explosion was complex, i.e.\ that the
spherically symmetric model described here must be an
oversimplification.

We turn now to the observed elemental abundance ratios in the summed \chandra\ spectrum
(Table~\ref{tab:summed98s}), which are enhanced by a factor of $\sim$ 3 - 30
over solar.  These heavy elements must have been synthesized in the
core evolution and explosion.

In core-collapse supernovae, a strong shock wave forms and propagates
outward through the star (e.g., Shigeyama \& Nomoto 1990).  Behind the
shock, materials are explosively burned into heavy elements and
accelerated to $\sim$5000~km~s$^{-1}$.  At the composition interface,
the expanding materials are decelerated by the collision with the
outer layers, which induces the Rayleigh-Taylor instability (e.g.,
Ebisuzaki, Nomoto, \& Nomoto 1988; Arnett et al.\ 1989).  Heavy
elements are then mixed into outer layers.  Such a mixing induced at
the Si/O, C/He, and He/H interfaces brings Fe-peak elements, as well
as Si, Mg, Ne, O, and C, into the H-rich envelope.  These heavy
elements in many Rayleigh-Taylor fingers are rather well-mixed through
the interaction between the fingers (e.g., Hachisu et al.\ 1992).  We
thus assume that, although the heavy elements mixed into the H-rich
envelope consists of only a fraction of the core materials, the
abundance ratios among the mixed heavy elements are the same as in the
integrated core materials.  The mixing associated with the
Rayleigh-Taylor instabilities continues until the shock wave reaches
the surface and the expansion becomes homologous.

When the ejecta collides with the circumstellar matter, forward and
reverse shock waves form and the latter propagates back into the
H-rich envelope.  The mass of the reverse-shocked H-rich envelope
$\delta M_{\rm H}$ at the time of the observations depends on the
strength of the circumstellar interaction and the density structure of
the H-rich envelope (e.g., Chevalier \& Fransson 1994; Suzuki \&
Nomoto 1995).  We denote by $\delta M_{\rm Z}$ the mass of heavy
elements mixed into the reverse-shocked H-rich envelope.  The observed
mass fraction of heavy elements is then $\delta M_{\rm Z}$/$X \delta
M_{\rm H}$, where $X$ denotes the H mass fraction.

We briefly summarize here the three types of heavy elements ejected from
core-collapse SNe (e.g., Thielemann, Nomoto, \& Hashimoto 1996):
\begin{description}
\item[O, Ne, and Mg:] These elements are synthesized mostly before
collapse during core- and shell-C burning; some Ne is processed into
Mg and O during Ne burning.  In the explosion, these elements are
partially produced by explosive C burning, Ne is partially burned into
Mg and Si, and O and Mg are partially burned into Si.  The masses of
these elements are sensitive to the progenitor mass, $M$, increasing
with $M$.  However, the degree of dependence on $M$ varies across
elements.  For example, the Ne/Mg ratio depends on the conditions of C
and Ne burning, thus depending on $M$.
\item[Si, S, and Ar:] These elements are produced by explosive O burning
during the explosion.  Their masses do not much depend on $M$ since the
pre-collapse core structure is not very sensitive to $M$.
\item[Fe:] The mass of ejected Fe is very sensitive to the mass cut that
divides the compact remnant and the ejecta, or the amount of fall-back
matter.  Theoretically, it is not possible to make a good estimate of
the Fe mass because of the uncertainties in the explosion mechanism.
\end{description}

In our comparisons, we use Si for normalization because it is the
least dependent on $M$.  We do not use Fe for comparison because of
the large theoretical uncertainties.  We now compare the O/Si, Ne/Si,
and Mg/Si ratios (relative to the solar abundance ratios) with the
theoretical calculations of \citet{nom97}.  These models are
summarized in Table~\ref{tab:nomoto}.
 
The observed O/Si value of 0--1.2 seems to indicate that the mass of
the progenitor was $\la$20~\Msun. The Mg/Si value of 0--0.7 also points
to a progenitor mass of $<$20~\Msun.  The observed Ne/Si ratio of
0.6--14 excludes progenitor masses of $\la$15~\Msun.  According to
these models, the mass of the proginitor must have been between
15~\Msun\ and 20~\Msun.

We note that the supernova yields are determined by the helium core
mass $M_{\rm He}$ of the progenitor rather than the zero-age
main-sequence mass $M$.  Thus the Mg/Si and Ne/Si results actually
provide the constraint $M_{\rm He} = 5 \pm 1~\Msun$ \citep{nom97}.
The $M$-$M_{\rm He}$ relation depends on the mass loss.  If we adopt
the spiral-in binary scenario for the origin of Type IIn SN
\citep{nomoto95}, a large amount of mass-loss in the early phase of
the progenitor's expansion leads to a smaller $M_{\rm He}$ compared
with the case of no mass loss.  In this case, the progenitor of
SN~1998S could be initially more massive.

\section{Summary}
Observations at radio, optical, and X-ray wavelengths have allowed the
estimation of the physical parameters of two quite different Type II
SN examples, Type II-P SN~1999em and Type IIn SN~1998S.  From these
results, it is possible to study, over a broad wavelength range, the
physical parameters of the explosions and the structure and density of
the CSM established by the pre-SN wind.  Such results are of
importance for estimation of the properties of the pre-SN star and its
last stages of evolution before explosion (e.g., \citealt{nomoto95}).
In addition, the \chandra\ data have revealed the presence of many
heavy elements in the spectrum of SN~1998S.  We have shown the kind of
analysis that can be done by comparing the observed X-ray determined
abundance ratios to the model predictions, and we were able to show
that the mass of the progenitor was $\sim$15--20~\Msun. We do caution,
however, that our model assumptions on mixing are approximate and that
there are uncertainties in both the interpretation of the X-ray data
and the model calculations. Because of the great diversity of Type II
(and related Type Ib/c) SNe, only the multiwavelength study of
additional examples can yield a comprehensive understanding of the
last stages of massive star evolution.

\section{Acknowledgments}
DP acknowledges that this material is based upon work supported under
a National Science Foundation Graduate Fellowship. WHGL and RAC gratefully
acknowledge support from NASA.  RAC is grateful to E.\ Baron for useful
correspondence.  KWW wishes to thank the Office of
Naval Research (ONR) for the 6.1 funding supporting this research.
Additional information and data on radio emission from SNe can be
found on \url{http://rsd-www.nrl.navy.mil/7214/weiler/} and linked
pages.  AVF is grateful for the support of NASA/ {\it Chandra} grants GO0-1001C and
GO1-2062D.  KN has been supported in part by the Grant-in-Aid for COE
Scientific Research (07CE2002, 12640233) of the Ministry of Education,
Science, Culture, and Sports in Japan.

\clearpage

\begin{deluxetable}{cccccc}
\tablewidth{0pt}
\tablecaption{\chandra\ Observations of SN~1999em. \label{tab:obs99em-xray}}
\tablecolumns{6}
\tablehead{
\colhead{Date} & \colhead{Day} & \colhead{Length} &
\colhead{Length After Filtering} & \colhead{Count Rate} & 
\colhead{\Fx[2--10~keV]/} \\
\colhead{} & \colhead{(From Ref.)} & \colhead{(ksec)} & 
\colhead{(ksec)} & \colhead{(cts ksec$^{-1}$)} & \colhead{\Fx[0.4--2~keV]}
}
\startdata
1999 Nov 1&   4&  	29.0&  21.6&   3.70& $2.1\pm 0.9$\\
1999 Nov 13&   16&       26.0&  10.6&   2.46& $0.8\pm 0.4$\\
1999 Dec 16&   49&       35.3&  35.1&   1.02& $0.7\pm 0.5$\\
2000 Feb 7&   102&      38.5&  23.9&   1.84& $0.2\pm 0.1$\\
2000 Oct 30&   368&      26.4&  26.4&   0.46& $0.9\pm 0.5$\\
2001 Mar 9&  495&       26.7&   24.6&   0.31& $0$\tablenotemark{*}\\
2001 Jul 22& 630&       29.8&   25.1&   0.19& $0$\tablenotemark{*}\\
\enddata
\tablenotetext{*}{No flux detected above 2~keV.}
\end{deluxetable}

\begin{deluxetable}{ccccc}
\tablewidth{0pt}
\tablecaption{\chandra\ Observations of SN~1998S. \label{tab:obs98s-xray}}
\tablecolumns{5}
\tablehead{
\colhead{Date}& \colhead{Day}& \colhead{Length}&
\colhead{Length After Filtering} & \colhead{Count Rate} \\
\colhead{}& \colhead{(From Ref.)}& \colhead{(ksec)}& \colhead{(ksec)}& \colhead{(cts ksec$^{-1}$)}}
\startdata
2000 Jan 10&  678&	18.9& 18.9&	34.4\\
2000 Mar 7&  735&	23.4& 23.1&	30.3\\
2000 Aug 1&  882&	19.8& 19.8&	25.8\\
2001 Jan 14&  1048&	29.2& 29.2& 	19.5\\
2001 Oct 17&  1324&     28.7& 14.8&     12.4\\
\enddata
\end{deluxetable}

\begin{deluxetable}{ccccc}
\tablewidth{0pt}
\tablecaption{X-ray Temperature Evolution of SN~1999em. \label{tab:fits99em}}
\tablecolumns{5}
\tablehead{
\colhead{Day}& \colhead{Obs. \mekal\ $kT$}&
\colhead{Obs. Brems. $kT$}& \colhead{Model $kT$ ($n=7$)}&
\colhead{Model $kT$ ($n=9$)}\\
\colhead{(From Ref.)}& \colhead{(keV)}& \colhead{(keV)}&
\colhead{(keV)}& \colhead{(keV)}
}
\startdata
4&    $5.0^{+5.1}_{-2.1}$&   $5.2^{+11.5}_{-2.3}$&   9.2& 4.1\\
16&   $2.5^{+2.7}_{-0.9}$&   $2.3^{+5.9}_{-1.1}$&   5.4& 2.8\\
49&   $2.9^{+2.6}_{-1.1}$&   $2.2^{+3.2}_{-0.9}$&   3.4& 2.0\\
102&  $0.8^{+0.2}_{-0.1}$&   $1.1^{+0.7}_{-0.4}$&   2.5& 1.6\\
368&  $1.0^{+0.7}_{-0.3}$&   $1.5^{+6.5}_{-0.8}$&   1.5& 1.1\\
\enddata
\tablecomments{Listed uncertainties are 90\% confidence intervals.}
\end{deluxetable}

\begin{deluxetable}{ccccc}
\tablewidth{0pt}
\tablecaption{\vmek\ Fits to the Spectra of SN~1998S. \label{tab:fits98s}}
\tablecolumns{5}
\tablehead{
\colhead{Day}& \colhead{$kT$}&
\colhead{Fe Abund.}&
\colhead{$\chi^2$/d.o.f.} &\colhead{\Lx [2--10~keV]\tablenotemark{a}} \\
\colhead{(From Ref.)}& \colhead{(keV)}& \colhead{(w.r.t. solar)}
& & \colhead{(\ergsec)}}
\startdata
678&    $10.4_{-2.2}^{+81.9}$& $6.5_{-6.5}^{+63}$&  82.7/96&  $9.3\times 10^{39}$\\
735&    $9.6_{-2.2}^{+5.5}$&   $4.2_{-2.6}^{+7.3}$& 71.7/71&  $8.8\times 10^{39}$\\
882&    $10.4_{-4.2}^{+14.1}$& $2.8_{-2.8}^{+15.8}$&65.8/75&  $6.1\times 10^{39}$\\
1048&    $8.0_{-2.1}^{+3.7}$&   $5.4_{-3.5}^{+14.5}$&57.0/59&  $5.3\times 10^{39}$\\
\enddata
\tablecomments{Listed uncertainties are 90\% confidence intervals.}
\tablenotetext{a}{for $z=0.003$ and $H_0=50$}
\end{deluxetable}

\begin{deluxetable}{lll}
\tablewidth{0pt}
\tablecaption{Elemental Abundances from \vmek\ Fit to Summed Spectrum of
SN~1998S. \label{tab:summed98s}}
\tablecolumns{4}
\tablehead{
\colhead{Element}& \colhead{Best-fit Abund.}& \colhead{90\% Confidence Interval}
}
\startdata
O& 0.7& 0--2.9\\
Ne& 15& 8.7--35\\
Mg& 0& 0--1.6\\
Al& 33& 0--135\\
Si& 5.7& 2.4--15\\
S& 8.7& 3.0--24\\
Ar& 18& 3.4--52\\
Ca& 0.8& 0--13\\
Fe& 3.0& 1.8--6.8\\ \hline
C& 1.0 & 0--200\\
N& 0 & 0--18\\
Na& 0& 0--36\\ 
Ni& 0& 0--14\\
\enddata
\tablecomments{These abundances are actually the ratio of a given
element to H normalized to the similar solar ratio.  For example,
``Ne'' is actually [(Ne/H)$_{\rm 98S}$]/[(Ne/H)$_\odot$].}
\end{deluxetable}

\begin{deluxetable}{lccccccc}
\tablewidth{0pt}
\tablecaption{Elemental Abundance Ratios Relative to Solar for SN~1998S and Theoretical
Models. \label{tab:nomoto}}
\tablecolumns{8}
\tablehead{
\colhead{Abund.\ ratio}& \colhead{SN~1998S}& \colhead{13~\Msun}& 
\colhead{15~\Msun}& \colhead{18~\Msun}& \colhead{20~\Msun}&
\colhead{25~\Msun}& \colhead{40~\Msun}
}
\startdata
 Ne/Si    &    0.6--14 &   0.14&   0.17&   0.86&   1.1&   2.2&   0.56\\
 Mg/Si    &    0--0.7   & 0.18  & 0.49  & 0.58  & 1.09  & 2.1  & 1.1\\
 O/Si	  &    0--1.2    & 0.20  &  0.43   &  0.80  &  1.4  &  2.4 & 1.6\\
\enddata
\end{deluxetable}


\begin{thebibliography}{}
\bibitem[Anders \& Grevesse(1989)]{andgrev89}Anders, E., \& Grevesse,
N.\ 1989, \gca, 53\ 197

\bibitem[Arnaud(1996)]{arnaud96}Arnaud, K.\ 1996, in Jacoby, G., \&
Branes, J., eds, ASP Conf.\ Ser.\ Vol.\ 101, Astronomical Data Analysis
Software and Systems V, Astron.\ Soc.\ Pac., San Francisco

\bibitem[Arnet et al.(1989)]{arn89}Arnett, W.\ D., Bahcall, J.\ N.,
Kirshner, R., \& Woosley, S.\ E.\ 1989, \araa, 27, 629

\bibitem[Baron et al.(2000)]{baron00}Baron, E., Branch, D.,
Hauschildt, P.\ H., Filippenko, A.\ V., Kirshner, R.\ P., Challis, P.\ M.,
Jha, S., Chevalier, R.\ A., Fransson, C., Lundqvist, P., Garnavich, P.,
Leibundgut, B., McCray, R., Michael, E., Panagia, N., Phillips, M.\ M.,
Pun, C.\ S.\ J., Schmidt, B., Sonneborn, G., Suntzeff, N.\ B., Wang,
L., \& Wheeler, J.\ C.\ 2000, \apj, 545, 444


\bibitem[Bloom et al.(1999)]{bloom99}Bloom, J.\ S., Kulkarni, S.\ R.,
Djorgovski, S.\ G., Eichelberger, A.\ C., Cote, P., Blakeslee, J.\ P.,
Odewahn, S.\ C., Harrison, F.\ A., Frail, D.\ A., Filippenko, A.\ V.,
Leonard, D.\ C., Riess, A.\ G., Spinrad, H., Stern, D., Bunker, A., Dey,
A., Grossan, B., Perlmutter, S., Knop, R.\ A., Hook, I.\ M., \& Feroci,
M.\ 1999, \nat, 401, 453


\bibitem[Branch et al.(1981)]{branch81}Branch, D., Falk, S.\ W.,
McCall, M.\ L., Rybski, P., Uomoto, A.\ K., \& Wills, B.\ J.\ 1981, \apj, 244, 780

\bibitem[Bregman \& Pildis(1992)]{bregpil92}Bregman, J. N.,\& Pildis,
R. A.\ 1992, \apj, 398, L107 

\bibitem[Burrows et al.(2000)]{bur00}Burrows, D., Michael, E., Hwang,
R., McCray, R., Chevalier, R., Petre, R., Garmire, G., Gordon, P.,
Holt, S., Nousek, J.\ 2000, \apj, 543, L149

\bibitem[Canizares, Kriss, \& Feigelson(1982)]{can82}Canizares, C.\
R., Kriss, G.\ A., \& Feigelson, E.\ D.\ 1982, \apj, 253, L17

\bibitem[Cash(1979)]{cash79}Cash, W.\ 1979, \apj, 228, 939

\bibitem[Chevalier(1976)]{chev76}Chevalier, R.\ A.\ 1976, \apj\ 207, 872

\bibitem[Chevalier(1982)]{chev82}Chevalier, R.\ A.\ 1982, \apj, 259, 302


\bibitem[Chevalier \& Fransson(1994)]{chevfran94}Chevalier, R.\ A., \&
Fransson, C.\ 1994, \apj, 420, 268

\bibitem[Chevalier(1998)]{chev98}Chevalier, R.\ A.\ 1998, \apj, 499, 810

\bibitem[Chugai(2001)]{chug01}Chugai, N.\ N.\ 2001, \mnras, submitted (astro-ph/0106234)

\bibitem[Chugai(1993)]{chug93}Chugai, N.\ N.\ 1993, \apj, 414, L101

\bibitem[Chugai \& Danziger(1994)]{chug94}Chugai, N.\ N., \& Danziger,
I.\ J.\ 1994, \mnras, 268, 173

\bibitem[LRISp; Cohen(1996)]{cohen96}Cohen, M.\ H.\ 1996, The LRIS Polarimeter
(Keck Obs. Instrument Manual), \url{http://www2.keck.hawaii.edu:3636/}


\bibitem[Eastman et al.(1994)]{eastman94}Eastman, R. G., Woosley,
S. E., Weaver, T. A., \& Pinto, P. A.\ 1994, \apj, 430, 300

\bibitem[Dennerl et al.(2001)]{dennerl01}Dennerl, K., et al., 2001,
\aap, 365, L202

\bibitem[Ebisuzaki, Shigeyama, \& Nomoto(1989)]{ebi89}Ebisuzaki, T.,
Shigeyama, T., \& Nomoto, K.\ 1989, \apj, 344, L65

\bibitem[Fabian \& Terlevich(1996)]{fab96}Fabian, A.\ C.,\&
Terlevich, R.\ 1996, \mnras, 280, L5 

\bibitem[Fassia et al.(2000)]{fas00}Fassia, A., Meikle, W.\ P.\ S.,
Vacca, W.\ D., Kemp, S.\ N., Walton, N.\ A., Pollacco, D.\ L., Smartt, S.,
Oscoz, A., Arag\'{o}n-Salamanca, A., Bennett, S., Hawarden, T.\ G.,
Alonso, A., Alcalde, D., Pedrosa, A., Telting, J., Arevalo, M.\ J.,
Deeg, H.\ J., Garz\'{o}n, F., G\'{o}mez-Rold\'{a}n, A., G\'{o}mez, G.,
Guti\'{e}rrez, C., L\'{o}pez, S., Rozas, M., Serra-Ricart, M., \&
Zapatero-Osorio, M.\ R.\ 2000, \mnras, 318, 1093

\bibitem[Fesen \& Becker(1990)]{fesbeck90}Fesen, R.\ A., \& Becker,
R.\ H.\ 1990, \apj, 351, 437

\bibitem[Fesen \& Matonick(1993)]{fessmat93}Fesen, R.\ A., \& Matonick,
D.\ M.\ 1993, \apj, 407, 110


\bibitem[Filippenko \& Barth(1997)]{filippenko97} Filippenko, A.\ V.,
\& Barth, A.\ J.\ 1997, \iaucirc, 6794  
 

\bibitem[Filippenko(1997)]{fil97}Filippenko, A.\ V.\ 1997, \araa, 35, 309

\bibitem[Filippenko \& Moran(1998)]{filmor98}Filippenko, A.\ V., \& Moran,
E. C.\ 1998, \iaucirc, 6830

\bibitem[Filippenko et al.(2001)]{fil01}Filippenko, A.\ V., et
al. 2001, in preparation

\bibitem[Fox et al.(1999)]{fox99}Fox, D.\ W., \& Lewin, W.\ H.\ G.\ 1999, \iaucirc,
7318

\bibitem[Fox et al.(2000)]{fox00}Fox, D.\ W., Lewin, W.\ H.\ G., Fabian,
A., Iwasawa, K., Terlevich, R., Zimmermann, H.\ U., Aschenbach, B.,
Weiler, K., Van Dyk, S., Chevalier, R., Rutledge, R., Inoue, H., \&
Uno, S.\ 2000, \mnras, 319, 1154

\bibitem[Fransson, Lundqvist, \& Chevalier(1996; hereafter,
FLC96)]{flc96}Fransson, C., Lundqvist, P., \& Chevalier, R.\ A.\ 1996, \apj,
461, 993, (FLC96)

\bibitem[Fransson \& Bj\"{o}rnsson(1998)]{franbjor98}Fransson, C., \&
Bj\"{o}rnsson, C.-I.\ 1998, \apj, 509, 861

\bibitem[Garnavich, Jha, \& Kirshner(1998a)]{gar98}Garnavich, P., Jha, S., \&
Kirshner, R.\ 1998a, \iaucirc, 6832

\bibitem[Garnavich et al.(1998b)]{gar98b}Garnavich, P., Kirshner, R.,
Challis, P., Koranyi, D., \& Culkins, M.\ 1998b, \iaucirc, 6845

\bibitem[Gehrels(1986)]{gehrels86}Gehrels, N.\ 1986, \apj, 303, 336

\bibitem[Gorenstein, Hughes, \& Tucker(1994)]{gore94}Gorenstein, P.,
Hughes, J.\ P., Tucker, W.\ H., \apj, 420, L25

\bibitem[Granslo et al.(1998)]{gran98}Granslo, B.\ H., Shanklin, J.,
Carvajal, J., \& Hornoch, K.\ 1998, \iaucirc, 6846

\bibitem[Hachisu et al.(1992)]{hach92}Hachisu, I., Matsuda, T.,
Nomoto, K., \& Shigeyama, T.\ 1992, \apj, 390, 230

\bibitem[Hamuy et al.(2001)]{ham01}Hamuy, M., et al.\ 2001, \apj, in
press (astro-ph/0105006)

\bibitem[Hasinger, Aschenbach, \& Tr\"{u}umper(1996)]{hasinger96}Hasinger,
G., Aschenbach, B., \& Tr\"{u}umper, J.\ 1996, \aap, 312, L9

\bibitem[H\"{o}flich(1991)]{hof91}H\"{o}flich, P.\ 1991, A\&A, 246, 418

\bibitem[Houck et al.(1998)]{houck98}Houck, J.\ C., Bregman, J.\ N.,
Chevalier, R.\ A., \& Tomisaka, K.\ 1998, \apj, 493, 431

\bibitem[Hughes et al.(2000)]{hughes00}Hughes, J.\ P., Rakowski, C.\
E., Burrows, D.\ N., \& Slane, P.\ O.\ 2000, \apj, 528, L109

\bibitem[Immler, Pietsch, \& Aschenbach(1998a)]{immler98a}Immler, S.,
Pietsch, W., \& Aschenbach, B.\ 1998a, \aap, 331, 601

\bibitem[Immler, Pietsch, \& Aschenbach(1998b)]{immler98b}Immler, S.,
Pietsch, W., \& Aschenbach, B.\ 1998b, \aap, 336, L1

\bibitem[Immler, Aschenbach, \& Wang(2001)]{immler01}Immler, S.,
Aschenbach, B., Wang, Q.\ D.\  2001, \apj, 561, L107

\bibitem[Immler, Wilson, \& Terashima(2002)]{immler02a}Immler, S.,
Wilson, A.\ S., Terashima, Y.\ 2002, submitted to ApJ

\bibitem[Kaaret(2001)]{kaaret01}Kaaret, P.\ 2001, \apj, 560, 715

\bibitem[Khokhlov et al.(1999)]{kho99}Khokhlov, A.\ M., H\"{o}flich,
P.\ A., Oren, E.\ S., Wheeler, J.\ C., Wang, L., \& Chtchelkanova,
A.\ Yu.\ 1999, \apj, 524, L107


\bibitem[Lacey et al.(2002)]{lacey01}Lacey, C.\ K., Weiler, K.\ W.,
Van
Dyk, S.\ D., Panagia, N., \& Sramek, R.\ A.\ 2002, in preparation

\bibitem[Leibundgut et al.(1991)]{leib91}Leibundgut, B., Kirshner,
R.\ P., Pinto, P.\ A., Rupen, M.\ P., Smith, R.\ C., Gunn, J.\ E., \&
Schneider, D.\ P.\ 1991, \apj, 372, 531

\bibitem[Leonard et al.(2000)]{leon00}Leonard, D.\ C., Filippenko,
A.\ V., Barth, A.\ J., \& Matheson, T.\ 2000, \apj, 536, 239

\bibitem[Leonard et al.(2001a)]{leon01a}Leonard, D.\ C., Filippenko,
A.\ V., Ardila, D.\ R., \& Brotherton, M.\ S.\ 2001a, \apj, in press

\bibitem[Leonard et al.(2001b)]{leon01b}Leonard, D.\ C., et al.,
2001b, \pasp, submitted


\bibitem[Li et al.(1998)]{li98}Li, W.-D., Li, C., Filippenko, A.\ V., \& Moran,
E.\ C.\ 1998, \iaucirc~6829

\bibitem[Li et al.(1999)]{li99}Li, W.-D., et al.\ 1999, \iaucirc, 7294

\bibitem[Li et al.(2000)]{li00}Li, W. D., et al.\ 2000, in Cosmic
Explosions, ed.\ S.\ S.\ Holt \& W.\ W.\ Zhang, (New York: AIP), p. 103

\bibitem[Liu et al.(2000)]{liu00}Liu, Q.-Z., Hu, J.-Y., Hang, H.-R.,
Qiu, Y.-L., Zhu, Z.-X., \& Qiao, Q.-Y.\ 2000, A\&AS, 144, 219

\bibitem[Lundqvist \& Fransson(1991)]{lf91}Lundqvist, P., \& Fransson, C.,
1991, ApJ, 380, 575

\bibitem[Maeda et al.(2000)]{maeda00}Maeda, K., Nakamura, T., Nomoto,
K., Mazzali, P.\ A., Patat, F., \& Hachisu, I.\ 2000, \apj, submitted
(astro-ph/0011003)

\bibitem[Matzner \& McKee(1999)]{matz99}Matzner, C.\ D., \& McKee,
C.\ F.\ 1999, \apj, 510, 379

\bibitem[Mewe, Gronenschild, \& van den Oord(1985)]{mewe85}Mewe, R., Gronenschild,
E.\ H.\ B.\ M., \& van den Oord, G.\ H.\ J.\ 1985, A\&AS, 62\ 197

\bibitem[Miller \& Stone(1993)]{millstone93}Miller, J.\ S., \& Stone,
R.\ P.\ S.\ 1993, Lick Obs. Tech. Rep. 66


\bibitem[Niemela, Ruiz, \& Phillips(1985)]{niem85}Niemela, V.\ S., Ruiz, M.\ T., \&
Phillips, M.\ M.\ 1985, \apj, 289, 52

\bibitem[Nomoto, Iwamoto, \& Suzuki(1995)]{nomoto95}Nomoto, K.,
Iwamoto, K., \& Suzuki, T.\ 1995, Phys.\ Rep., 256, 173

\bibitem[Nomoto et al.(1997)]{nom97}Nomoto, K., Hashimoto, M.,
Tsujimoto, T., Thielemann, F.-K., Kishimoto, N., Kubo, Y., \&
Nakasato, N.\ 1997, Nucl.\ Phys.\ A, 616, 79 

\bibitem[Oke et al.(1995)]{oke95}Oke, J.\ B., Cohen, J.\ G., Carr, M.,
Cromer, J., Dingizian, A., Harris, F.\ H., Labrecque, S., Lucinio, R.,
Schaal, W., Epps, H., \& Miller, J.\ 1995, \pasp, 107, 375

\bibitem[Panagia et al.(1980)]{pan80}Panagia, N., Vettolani, G.,
Boksenberg, A., Ciatti, F., Ortolani, S., Rafanelli, P., Rosino, L.,
Gordon, C., Reimers, D., Hempe, K., Benvenuti, P., Clavel, J., Heck,
A., Penston, M.\ V., Macchetto, F., Stickland, D.\ J., Bergeron, J.,
Tarenghi, M., Marano, B., Palumbo, G.\ G.\ C., Parmar, A.\ N., Pollard,
G.\ S.\ W., Sanford, P.\ W., Sargent, W.\ L.\ W., Sramek, R.\ A., Weiler,
K.\ W., \& Matzik, P.\ 1980, \mnras\ 192, 861

\bibitem[cf.\ Patat et al.(1993)]{patat93}Patat, F., Barbon, R.,
Cappellaro, E., \& Turatto, M.\ 1993, \aap, 98, 443

\bibitem[cf.\ Patat et al.(1994)]{patat94}Patat, F., Barbon, R.,
Cappellaro, E., \& Turatto, M.\ 1994, \aap, 282, 731

\bibitem[Petre et al.(1994)]{petre94}Petre, R., Okada, K., Mihara, T.,
Makishima, K., \& Colbert, E.\ J.\ M.\ 1994, \pasj, 46, L115

\bibitem[Pian et al.(1999)]{pia99}Pian, E., Amati, L, Antonelli, L.,
Butler, R, Costa, E., Cusumano, G., Danziger, J., Feroci, M., Fiore,
F., Frontera, F., Giommi, P., Masetti, N., Muller, J., Oosterbroek,
T., Owens, A., Palazzi, E., Piro, L., Castro-Tirado, A., Coletta, A.,
dal Fiume, D., del Sordo, S., Heise, J., Nicastro, L., Orlandini, M.,
Parmar, A., Soffitta, P., Torroni, V., in't Zand, J.\ 1999, A\&AS,
138, 399

\bibitem[Pian et al.(2000)]{pian00}Pian, E, et al.\ 2000, \apj, 536, 778

\bibitem[Predehl \& Schmitt(1995)]{predschm95}Predehl, P., \& Schmitt,
J.\ H.\ M.\ M.\ 1995, \aap, 293, 889

\bibitem[Ray, Petre, \& Schlegel(2001)]{ray01}Ray, A., Petre, R.,
Schlegel, E.\ M.\ 2001, \aj, 122, 966



\bibitem[Schlegel(1990)]{schlegel90}Schlegel, E.\ M.\ 1990, \mnras, 244, 269

\bibitem[Schlegel(1994)]{schlegel94}Schlegel, E.\ M.\ 1994, \aj, 108, 1893

\bibitem[Schlegel(1995)]{schlegel95}Schlegel, E.\ M.\ 1995, Rep.\
Prog.\ Phys., 58, 1375

\bibitem[Schlegel, Petre, \& Colbert(1996)]{schlegel96}Schlegel, E.\ M.,
Petre, R., Colbert, E.\ J.\ M.\ 1996, \apj, 456, 187

\bibitem[Schlegel(1999a)]{schlegel99a}Schlegel, E.\ M., Ryder, S.,
Staveley-Smith, L., Petre, R., Colbert, E., Dopita, M., \&
Campbell-Wilson, D.\ 1999a, \aj, 118, 2689

\bibitem[Schlegel(1999b)]{schlegel99b}Schlegel, E.\ M.\ 1999b, \apj, 527, L85

\bibitem[Schlegel(2001a)]{schlegel01a}Schlegel, E.\ M.\ 2001a, \apj, 556, L25

\bibitem[Schlegel(2001b)]{schlegel01b}Schlegel, E.\ M.\ 2001b, in preparation

\bibitem[Shigeyama \& Nomoto(1990)]{shigno90}Shigeyama, T., \& Nomoto,
K.\ 1990, \apj, 360, 242

\bibitem[Smartt et al.(2001)]{smartt01}Smartt, S.\ J., Gilmore, G., F.,
Tout, C.\ A., \& Hodgkin, S.\ T.\ 2001, \apj, in press (astro-ph/0107499)

\bibitem[Sohn \& Davidge(1998)]{sohndav98}Sohn, Y.-J., \& Davidge, T.\
J.\ 1998, \aj, 115, 130

\bibitem[Sollerman, Cumming, \& Lundqvist(1998)]{sol98}Sollerman, J., Cumming,
R.\ J., \& Lundqvist, P.\ 1998, \apj, 493, 933


\bibitem[Suzuki \& Nomoto(1995)]{suzno95}Suzuki, T., \& Nomoto,
K.\ 1995, \apj, 455, 658

\bibitem[Thielemann, Nomoto, \& Hashimoto(1996)]{thiel96}Thielemann,
F.-K., Nomoto, K., \& Hashimoto, M.\ 1996, \apj, 460, 468

\bibitem[Tully(1988)]{tully88}Tully, R.\ B.\ 1988, {\it Nearby Galaxies
Catalog}, Cambridge University Press

\bibitem[Van Dyk et al.(1993)]{vandyk93}Van Dyk, S.\ D., Weiler, K.\ W.,
Sramek, R.\ A., \& Panagia, N.\ 1993, \apj, 419, L69



\bibitem[Van Dyk et al.(1999)]{vandyk99}Van Dyk, S.\ D., Lacey, C.\ K.,
Sramek, R.\ A., \& Weiler, K.\ W.\ 1999, \iaucirc, 7322

\bibitem[Wang et al.(2001)]{wang01}Wang, L., Howell, D.\ A.,
H\"{o}flich, P., \& Wheeler, J.\ C.\ 2001, \apj, in press

\bibitem[Weiler et al.(1986)]{weiler86}Weiler, K.\ W., Sramek, R.\ A.,
Panagia, N., van der Hulst, J.\ M., \& Salvati, M.\ 1986, \apj, 301, 790

\bibitem[Weiler et al.(1990)]{weiler90}Weiler, K.\ W., Panagia, N., \&
Sramek, R.\ A.\ 1990, \apj, 364, 611

\bibitem[Weiler et al.(1991)]{weiler91}Weiler, K.\ W., Van Dky, S.\ D.,
Panagia, N., Sramek, R.\ A., \& Discenna, J.\ L.\ 1991, \apj, 380, 161

\bibitem[Weiler et al.(1992)]{weiler92}Weiler, K.\ W., Van Dyk, S.\ D.,
Panagia, N., \& Sramek, R.\ A.\ 1992, \apj, 398, 248

\bibitem[Wheeler et al.(2000)]{whee00}Wheeler, J.\ C., Yi, I.,
H\"{o}flich, P., \& Wang, L.\ 2000, \apj, 537, 810

\bibitem[Willick et al.(1997)]{willick97}Willick, J.\ A., Courteau, S.,
Faber, S.\ M., Burstein, D., Dekel, A., \& Strauss, M.\ A.\ 1997, \apj,
109, 333

\bibitem[Zimmermann et al.(1994)]{zim94}Zimmermann, U., Lewin, W.,
Predehl, P., Aschenbach, B., Fabbiano, G., Hasinger, G., Lubin, L.,
Magnier, E., van Paradijs, J., Petre, R., Pietsch, W., Tr\"{u}mper,
J.\ 1994, \nat, 367, 621
\end{thebibliography}
\end{document}